\documentclass[11pt]{article}
\usepackage[T1]{fontenc}
\usepackage{times}
\usepackage[scaled=.95]{helvet}
\usepackage[a4paper,textwidth=162mm,textheight=240mm,centering]{geometry}
\usepackage{graphicx,booktabs,xcolor,caption,rotating}
\usepackage{amsfonts,amssymb,amsbsy,amsmath,amsthm}
\usepackage[super,sort&compress,comma]{natbib}
\usepackage{array,longtable,needspace,tikz,url}
\usetikzlibrary{arrows.meta}
\definecolor{flowblue}{HTML}{0072B2}

\begin{document}
{\noindent\sffamily\bfseries\fontsize{18}{21}\selectfont Sandwich-adjusted Bayesian copula inference and model checks for joint laboratory extremes in emergency care\par}
\medskip
{\noindent\large Agnideep Aich$^{1}$, Md Monzur Murshed$^{2}$, Sameera Hewage$^{3}$, Ashit Baran Aich$^{4}$, Carl Preiksaitis$^{1,\dagger}$ and Christian Rose$^{1,\dagger}$\par}
\smallskip
{\noindent\small $^{1}$Department of Emergency Medicine, Stanford University School of Medicine, Palo Alto, CA, USA\\
$^{2}$Department of Mathematics and Statistics, Minnesota State University, Mankato, MN, USA\\
$^{3}$Department of Mathematics, Southern Utah University, Cedar City, UT, USA\\
$^{4}$Department of Statistics, formerly of Presidency College, Kolkata, India\\
$^{\dagger}$Carl Preiksaitis and Christian Rose share senior authorship.\par}
\medskip
\begin{abstract}
We estimate how often creatinine and aspartate aminotransferase are both high in the same emergency department specimen. We fit a rank-based Bayesian copula model, adjust its intervals for estimated margins and patient clustering, and compare fitted and observed joint rates at the recorded cut-offs. In simulations that fitted the generating family, 95\% sandwich-adjusted and patient-bootstrap intervals achieved 94.0--95.7\% coverage. In a retrospective cross-sectional analysis of electronic health records for 107,387 adults at a US emergency department in 2020--2023, both measurements exceeded their 95th percentiles 1.63 times as often as under independence (95\% confidence interval 1.49--1.77). Among eight families, the Student $t$ copula had the lowest pseudo-Bayesian information criterion but overestimated the observed 95th-percentile joint rate by 33.3\%. In equally large rounded samples generated from the fitted BB1 and BB7 copulas, the criterion selected Student $t$ from six candidate families in 98.1\% and 99.85\% of replicates, respectively. The paired bootstrap flagged the selected family's 95th-percentile discrepancy in 83.6\% and 81.9\%. Under independence, the false-flag rate reached 23.3\% at the 95th percentile with 512 rounded, clustered observations. Overall model rankings can conceal errors at the cut-offs of interest, and sparse joint events limit the interpretation of the check.
\end{abstract}

\par\medskip
\section{Introduction}\label{sec:introduction}

Creatinine and aspartate aminotransferase (AST) elevations can accompany the same acute illness, and in emergency care both are often measured on the same blood specimen. We estimate how often both measurements are high in the same specimen, and examine the coverage of intervals for the corresponding probability under a fitted copula. Three features of routine laboratory data can affect that coverage: the marginal distributions are estimated from the same observations, the measurements are rounded, and repeat visits from the same patient may be dependent. Our application uses paired creatinine and AST measurements from a single emergency department. The primary analysis uses each patient's first qualifying encounter, and repeat encounters enter only a secondary analysis of all encounters.

Joint elevations arise in several clinical settings. AST may reflect skeletal muscle injury in rhabdomyolysis.\citep{weibrecht2010} In hypoxic hepatitis, marked aminotransferase elevations are characteristic\citep{tapper2015} and often occur alongside acute kidney injury.\citep{drolz2016} Sepsis can involve dysfunction of several organs,\citep{singer2016} and hepatorenal syndrome, a complication of cirrhosis, is a functional kidney impairment.\citep{simonetto2020} Peak AST was positively associated with peak creatinine after liver transplantation.\citep{leithead2013} In patients undergoing primary percutaneous coronary intervention for ST-elevation myocardial infarction, Park et al.\citep{park2016} reported poorer outcomes among those with admission AST at least twice the upper normal limit and acute kidney injury defined by a subsequent creatinine rise within 48 hours after the procedure, compared with patients with either abnormality alone or neither. The co-occurrence of the two elevations therefore does not identify a single clinical process, and these associations come from selected patient groups, with some studies using peak values or measurements taken at different times. They do not directly describe how often both markers are high in the same emergency department specimen.

We address that question using emergency department data from 2020--2023, restricted to adults tested for both markers. The same-specimen measurements provide an observed joint rate. They also allow us to ask whether a model fitted to the dependence between the measurements reproduces that rate. Figure~\ref{fig:data-deciles} motivates this check: along the diagonal, the excess co-occurrence is largest where both measurements are in their highest or lowest groups, and some off-diagonal combinations occur less often than expected under independence.

Copulas model this dependence separately from each measurement's distribution.\citep{sklar1959,nelsen2006,genestfavre2007} A correlation coefficient alone does not determine how often both measurements exceed high cut-offs.\citep{joe2014} Copula families impose different assumptions about this joint behaviour. Gaussian and Student $t$ copulas with the same correlation parameter can give different probabilities of simultaneous high values.\citep{demarta2005} Other families allow dependence to differ between high and low values.\citep{brechmann2013} Hospital research has examined extreme counts of patients with flu-like symptoms and joint models for mortality and length of stay.\citep{ranjbar2022,marra2026}

In medical applications, the quantity being estimated determines which features of dependence matter. Berger et al.\ modelled the ratio of two correlated biomarkers,\citep{berger2025} and Escarela et al.\ used residuals to assess candidate copulas for a marker and a survival time.\citep{escarela2023} In paired diagnostic-test studies, Nikoloulopoulos allowed tail dependence and asymmetry when modelling between-study variation.\citep{nikoloulopoulos2024} McMenamin et al.\ estimated the probability of meeting several response criteria by jointly modelling their components.\citep{mcmenamin2021} Here the quantity is the probability that two recorded measurements both exceed their cut-offs. We can count that event in the same patients used to fit the model.

The relevant probabilities in our study are defined at specified percentile levels. Limiting tail dependence describes how the two variables behave together as thresholds approach the upper ends of the distributions.\citep{ledford1996,coles1999,schmidt2006,joe2010} We consider joint exceedances at the 90th and 95th percentiles and compare the rates with those expected under independence. These cut-offs describe high values within the tested population and do not define organ injury. The joint rates can be estimated directly from the recorded measurements. Obtaining the corresponding probabilities from a copula also requires estimates of each measurement's distribution.

Rank-based pseudo-likelihood meets this requirement without imposing parametric distributions on the measurements.\citep{genest1995} Extended rank likelihood offers another route and accommodates discrete observations.\citep{hoff2007} For uncertainty estimation, Bayesian copula methods provide distributions for dependence parameters and functions of them.\citep{smith2013} Such analyses involve choices about the copula family and prior specification.\citep{huard2006,guillotte2012,battagliese2023} Approximate inference can also target dependence measures directly.\citep{grazian2017} For each of the six main copula families, a rank-based pseudo-likelihood is combined with a restricted Jeffreys prior to form an initial, or working, posterior. Its uncertainty calculation treats the estimated marginal distributions as fixed and the encounters as independent.

The marginal distributions are estimated from the same observations used to fit the copula. Rounding produces tied values that can affect inference about dependence.\citep{genestneslehova2007,li2020} In the all-encounter analysis, repeat visits from the same patient may be dependent. We account for this grouping by resampling patients and retaining their observations together.\citep{field2007} The broader issue is that posterior probability and repeated-sampling coverage need not agree under model misspecification.\citep{kleijn2012,muller2013}

We therefore apply a sandwich adjustment to account for estimated marginal distributions and repeated observations within patients. This builds on work on composite likelihoods and estimating functions.\citep{ribatet2012,shaby2014} The adjustment accounts for uncertainty in the estimated copula probabilities. Whether those probabilities agree with the observed joint rates still requires a model check. General goodness-of-fit procedures examine agreement across the joint distribution.\citep{genest2009} Our comparison focuses on the laboratory cut-offs of interest and uses the same recorded cut-offs for fitted and empirical rates. Rounding can place many observations exactly at a cut-off, so the proportions strictly above it must be calculated from the data.

Building on these established methods, this paper assesses whether calibrated copula inference adequately describes the joint laboratory events observed in emergency care. The empirical rates provide a reference for judging what the dependence model adds and where it fails. We make the following contributions.

\begin{enumerate}
\item Calibration of a rank-based working posterior using a sandwich adjustment that accounts for estimated marginal distributions and groups observations by patient.

\item Comparison of fitted and observed joint rates at the observed cut-offs. We estimate both rates in the same patient bootstrap samples to obtain an interval for their difference.

\item Assessment of interval coverage in simulations with continuous measurements and with rounded measurements and repeated observations. We also simulate from two copulas fitted to the application to assess how often the family selected by pseudo-BIC gives the closest joint rate and how often the paired check detects a discrepancy. We report Monte Carlo standard errors.\citep{morris2019}

\item Analysis of joint creatinine and AST extremes in same-specimen measurements from 107,387 adult patients at the study emergency department during 2020--2023. We report separate results by age, acuity and recorded sex.
\end{enumerate}

The simulations showed improved interval coverage after the sandwich adjustment, with coverage similar to that of the patient bootstrap. In the hospital data, creatinine and AST were jointly high more often than expected under independence. Most fitted copulas did not reproduce all the observed joint rates. Among the eight families fitted to the overall sample, the Student $t$ copula had the lowest pseudo-Bayesian information criterion and overestimated the joint rate at the observed 95th-percentile cut-offs by 33.3\%. The survival Clayton copula had the highest criterion and overestimated the same rate by 4.2\%. In the application-sized simulations, the same criterion usually selected Student $t$, although its fitted 95th-percentile rate was lower than the generating rate on average.

The remainder of the paper is organised as follows. Section~\ref{sec:data} describes the study population and laboratory measurements. Section~\ref{sec:methods} presents the joint-rate definitions, copula estimation, interval adjustment, model checks and simulation design. Sections~\ref{sec:simulation-results} and~\ref{sec:application} report the simulation results and the analysis of the emergency department data. Section~\ref{sec:discussion} discusses the findings, limitations and future directions.

\section{Data}\label{sec:data}
This retrospective study uses encounters at an adult emergency department in the United States during 2020--2023. The analysis uses electronic health record extracts from the institutional clinical data warehouse, obtained on 29 September 2026. The reporting checklist for the clinical application follows STROBE for cross-sectional studies and its RECORD extension for routinely collected health data.\citep{vonelm2007,benchimol2015}

Study size was determined by the available eligible records over the four-year period. The study population comprises patients with both creatinine and AST tests available. The two results must share an encounter, a nonblank specimen identifier and an identical collection time. Collection must occur between arrival and departure and within six hours of arrival. The units must match mg/dL for creatinine and U/L for AST exactly; we do not convert units.

We first resolve duplicate laboratory records and drop records with unresolved conflicts. For each encounter, the earliest specimen pair with verified or corrected results meeting the specimen, timing and unit rules is then selected. The encounter is excluded if more than one eligible specimen shares the earliest collection time, or if either selected result has a reporting-limit marker, conflicting numeric fields, or a non-numeric, non-finite or non-positive value. A later specimen is not substituted when the selected pair fails these checks. Supplementary Appendix~\ref{app:records} gives the full record-selection rules.

Qualifying encounters require a recorded age of at least 18 years. For each patient, the first qualifying encounter is selected, and the patient is included if their recorded age at that encounter is 18--100 years. Because recorded ages above 100 years frequently reflect placeholder values, we excluded these records, removing 119 patients from the first-encounter sample and 239 encounters from the repeated-encounter sample. The resulting primary sample contains 107,387 patients, and the repeated-encounter sample contains 202,118 qualifying encounters at ages 18--100 years among the same patients. Thus, the repeated-encounter sample is restricted to patients retained in the first-encounter sample, and every included encounter must have a recorded age of 18--100 years.

Of 285,024 encounters at the adult emergency department with arrival during 2020--2023, a recorded age of at least 18 years, a recorded departure at or after arrival and a unique, non-conflicting encounter record, 202,413 had a first same-specimen creatinine and AST pair meeting the specimen, timing, status and unit rules above, after excluding 261 encounters with more than one specimen at the earliest collection time. We then excluded 15 pairs with a reporting-limit marker and 41 with non-numeric or non-finite values, leaving 202,357 encounters from 107,506 patients before the age exclusion described above.

In the primary sample, the median creatinine was 0.81~mg/dL (interquartile range 0.66--1.00) and the median AST was 27~U/L (21--37). The sample includes 41,323, 30,946 and 35,118 patients aged 18--44, 45--64 and 65--100 years; 31,237, 72,781 and 2,911 patients with Emergency Severity Index 1--2, 3 and 4--5, and 458 with missing acuity; and 58,445 patients recorded as female, 48,919 as male and 23 as Unknown.

\begin{figure}[!htbp]
\centering
\includegraphics[width=0.75\linewidth]{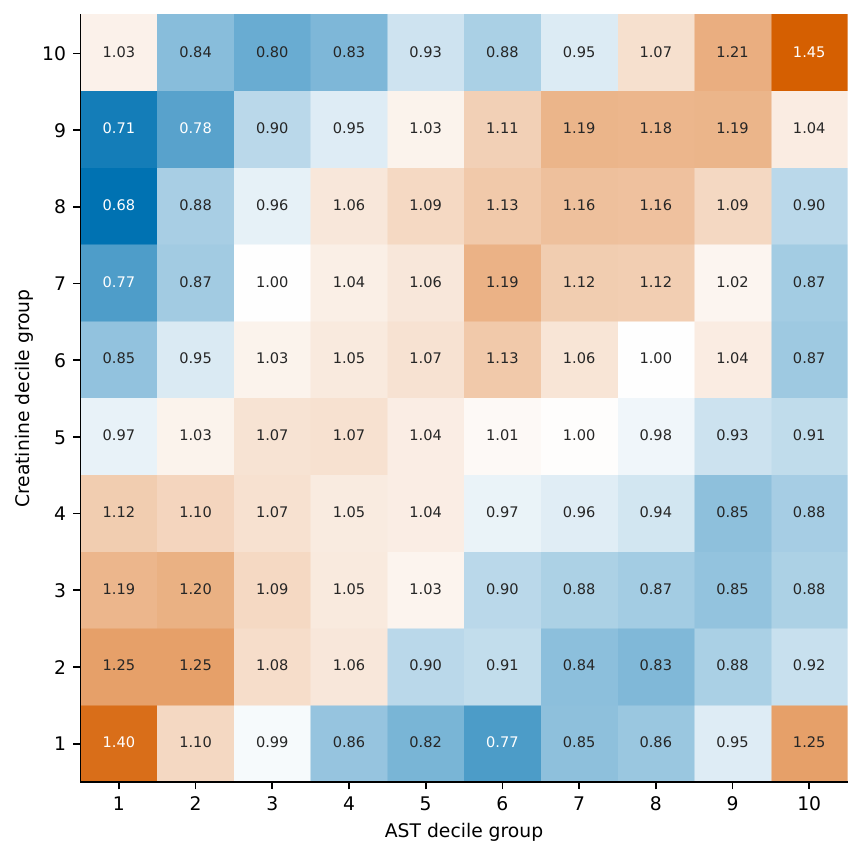}
\caption{Observed-to-independence ratios across creatinine and AST decile groups at the first qualifying encounter. Each cell gives the observed patient count divided by the count expected under independence.}\label{fig:data-deciles}
\par\smallskip
\begin{minipage}{0.95\linewidth}\footnotesize
\textbf{Alt text:} Rows show creatinine decile groups and columns show AST decile groups. All ten diagonal ratios exceed one. The ratio is 1.45 when both measurements are in their highest groups and 1.40 when both are in their lowest groups. The lowest creatinine group paired with the highest AST group has a ratio of 1.25. The two lowest ratios are 0.68 and 0.71, for creatinine groups 8 and 9, respectively, paired with the lowest AST group.
\end{minipage}
\end{figure}

Figure~\ref{fig:data-deciles} divides each measurement into decile groups and compares the number of patients in each pair of groups with the number expected under independence. Tied values stay together, so the groups need not contain equal numbers of patients; values above one indicate more patients than expected under independence. Every group on the diagonal has a ratio above one, ranging from 1.04 to 1.45, with the largest values where both measurements are in their highest group (1.45) or both in their lowest (1.40). The pattern is not the same in every corner: the lowest creatinine group combined with the highest AST group also has more patients than expected (1.25), whereas the highest creatinine group has fewer than expected alongside AST groups 2--4 (0.80--0.84). A single correlation coefficient does not summarise this pattern.

The institutional review board approved the study and waived informed consent for this retrospective study of existing electronic health records.

\section{Methods}\label{sec:methods}

This section defines the joint rates of interest and describes how the copulas are fitted, how uncertainty is quantified and how model fit is assessed. It then describes the group and sensitivity analyses and the simulation design. Supplementary Figure~\ref{fig:methods-flow} summarises these steps.

\subsection{Clinical question}
We study how often creatinine and AST are both unusually high in patients tested in the emergency department. An extreme value is an exceedance of a specified sample percentile. The observed frequency is compared with the frequency expected under independence, and the fitted copula models are assessed on whether they reproduce these probabilities.

Let $i=1,\ldots,N$ index the analysed encounters, with $X_i$ and $Y_i$ denoting the recorded creatinine and AST values. For each patient $m$, $\mathcal I_m$ is the set of indices identifying that patient's encounters, with $m=1,\ldots,M$. The primary analysis uses one encounter per patient, so $N=M$ and each set contains one index. In the repeated-visit analysis, every encounter has equal weight. Patients are treated as independent sampling units and are resampled to account for dependence between encounters from the same patient.

\subsection{Observed joint extremes}\label{sec:empirical}
For $q\in\{0.90,0.95\}$, let $x_q$ and $y_q$ be the inverse empirical distribution quantiles. Each is the smallest recorded value whose cumulative sample proportion is at least $q$. We define
\begin{equation}\label{eq:empirical}
\begin{split}
\widehat p_X(q)&=N^{-1}\sum_{i=1}^N\mathbf 1(X_i>x_q),\qquad
\widehat p_Y(q)=N^{-1}\sum_{i=1}^N\mathbf 1(Y_i>y_q),\\
\widehat p_{XY}(q)&=N^{-1}\sum_{i=1}^N\mathbf 1(X_i>x_q,\ Y_i>y_q),\\
\widehat K(q)&=\frac{\widehat p_{XY}(q)}{\widehat p_X(q)\widehat p_Y(q)}.
\end{split}
\end{equation}
The ratio $\widehat K(q)$ compares the observed joint rate with independence. A value of one corresponds to independence at these cut-offs. With rounded measurements, the proportion strictly above a cut-off can be smaller than $1-q$. The denominator is therefore the product of the observed marginal exceedance proportions. The cut-offs, joint counts, joint rates and ratios to independence are all reported.

Percentile confidence intervals are obtained by sampling patients with replacement and retaining all encounters from each selected patient. When a patient is selected more than once, their encounters are repeated accordingly. In every sample, we recalculate the cut-offs and all terms in equation~\eqref{eq:empirical}. The overall primary and all-encounter empirical intervals use 2,000 resamples. Group intervals, paired sensitivity comparisons and analyses involving copula refitting use 999 resamples (see Davison and Hinkley, Section 5.2).\citep{davison1997} All intervals are pointwise 95\% intervals.

\subsection{Rank-based copula estimation}\label{sec:copula}
To model dependence separately from the marginal distributions, we use copulas. For continuous margins, write
\begin{equation}\label{eq:copula}
\Pr(X\leq x,Y\leq y)=C_{\boldsymbol\theta}\{F_X(x),F_Y(y)\},\qquad
c_{\boldsymbol\theta}(u,v)=\frac{\partial^2 C_{\boldsymbol\theta}(u,v)}{\partial u\,\partial v}.
\end{equation}
The parameter $\boldsymbol\theta$ can have more than one component. We fit Gaussian, Student $t$, Frank, Gumbel, Joe and survival Clayton copulas. The Student $t$ copula has a dependence parameter and a degrees-of-freedom parameter, whereas the other five families have one parameter. Supplementary Appendix~\ref{app:families} gives their distribution functions and the parameter domains used in the analysis.

The marginal distributions are estimated with empirical ranks.\citep{genest1995} If $r_i$ and $s_i$ are the marginal mid-ranks, the pseudo-observations and log pseudo-likelihood are
\begin{equation}\label{eq:pseudolikelihood}
\widehat U_i=\frac{r_i}{N+1},\qquad
\widehat V_i=\frac{s_i}{N+1},\qquad
\ell_N(\boldsymbol\theta)=\sum_{i=1}^N\log c_{\boldsymbol\theta}(\widehat U_i,\widehat V_i).
\end{equation}
The copula parameters are estimated by maximising this expression over the specified parameter domain. The ranks are estimated from the same sample, and rounded values give discrete pseudo-observations. Equation~\eqref{eq:pseudolikelihood} defines the log pseudo-likelihood.

Ties arise when observations share the same recorded value. They are common in the primary sample, where 99.7\% of creatinine values and 99.7\% of AST values occur more than once. These ties can affect copula estimation.\citep{li2020} We use the continuous copula families as approximate models of dependence for the recorded values and assess sensitivity to tie handling by comparing mid-ranks with randomised ranks.

\subsection{Posterior distribution}\label{sec:posterior}
We combine the rank-based pseudo-likelihood with a restricted Jeffreys prior derived from the continuous copula model.\citep{jeffreys1946} Its information matrix includes both Student $t$ parameters. The resulting working posterior treats the estimated margins as known and the encounters as independent. Supplementary Appendix~\ref{app:numerics} gives the prior, posterior density and numerical integration checks.

To summarise joint extremes under the copula model, we define the nominal joint-tail probability for continuous uniform margins as
\begin{equation}\label{eq:nominal}
R_q(\boldsymbol\theta)=\Pr_{\boldsymbol\theta}(U>q,V>q)
=1-2q+C_{\boldsymbol\theta}(q,q).
\end{equation}
We then transform parameter draws through $R_q$ and report posterior means, medians and equal-tailed 95\% intervals. These summaries describe the nominal continuous-percentile probability in equation~\eqref{eq:nominal}, not the observed strict-exceedance probability in equation~\eqref{eq:empirical}. Both are finite-threshold quantities.

\subsection{Interval calibration}\label{sec:calibration}
We calibrate the posterior intervals by adjusting posterior uncertainty to target their stated coverage under repeated sampling. Following the sandwich approach of Shaby,\citep{shaby2014} we match the posterior covariance locally to the estimated sampling covariance. For our data, the covariance estimate accounts for empirical-margin estimation and patient clustering.

Let $\boldsymbol z$ denote the transformed parameter and $\widehat{\boldsymbol z}$ its estimate obtained by maximising the log pseudo-likelihood. The curvature $H_N=-\partial^2\ell_N/\partial\boldsymbol z\partial\boldsymbol z^{\mathsf T}$ is evaluated at this estimate. Let $\widehat{\boldsymbol\psi}_i$ denote the centred score contribution after the empirical-margin correction. Summing within patients gives $\widehat{\boldsymbol\Psi}_m=\sum_{i\in\mathcal I_m}\widehat{\boldsymbol\psi}_i$. We use
\begin{equation}\label{eq:sandwich}
\widehat V_z=H_N^{-1}
\left\{\frac{M}{M-1}\sum_{m=1}^M
\widehat{\boldsymbol\Psi}_m\widehat{\boldsymbol\Psi}_m^{\mathsf T}\right\}
H_N^{-\mathsf T}.
\end{equation}

We centre and rescale the draws in logit coordinates to respect the bounds. The corrected scores account for estimation of the margins and are summed within patients. Under the interior regularity conditions of Supplementary Proposition~\ref{prop:calibration}, the adjusted interval has first-order nominal coverage for the working-model probability. Under misspecification this probability can differ from the generating probability.\citep{chen2006,tsukahara2005} Supplementary Appendix~\ref{app:calibration} defines the target and gives the derivation.

For a frequentist comparison, we use a patient-cluster percentile bootstrap. In each resample, we rebuild the ranks and refit the copula. The 2.5th and 97.5th percentiles of the fitted probabilities in equation~\eqref{eq:nominal} form the confidence interval. In the primary analysis, sampling patients is the same as sampling rows. Boundary fits and computational failures are reported. A bootstrap interval requires all 999 refits to succeed, while a sandwich interval requires an interior estimate and an estimable covariance. No bootstrap refit failed in the application or in the reported simulation settings.

\subsection{Model fit}\label{sec:adequacy}
To assess whether fitted copulas reproduce the observed joint rates, we compare fitted and empirical probabilities at the same observed cut-offs. Set $a_q=\widehat F_X(x_q)=1-\widehat p_X(q)$ and $b_q=\widehat F_Y(y_q)=1-\widehat p_Y(q)$. The fitted probability and its difference from the observed rate are
\begin{equation}\label{eq:gap}
\begin{split}
\widehat p_C(q)&=1-a_q-b_q+C_{\widehat{\boldsymbol\theta}}(a_q,b_q),\\
\widehat\Delta(q)&=\widehat p_C(q)-\widehat p_{XY}(q).
\end{split}
\end{equation}
The values $a_q$ and $b_q$ account for the observed marginal proportions above rounded cut-offs. In each patient resample, we recalculate the cut-offs, marginal proportions, ranks and fitted parameters, then calculate $\widehat\Delta(q)$ from that same resample. Its percentile interval includes variation in both the empirical rate and the fitted model. An interval excluding zero indicates a discrepancy at that threshold. This is a pointwise tail check, not an overall goodness-of-fit test.

All fitted families are reported. For descriptive comparison, we use the pseudo-Bayesian information criterion (pseudo-BIC) $-2\ell_N(\widehat{\boldsymbol\theta})+d\log N$, where $d$ is the number of fitted parameters. These comparisons use the same sample and rank representation. For repeated encounters, $N$ is the encounter count.

\subsection{Groups and sensitivity analyses}\label{sec:sensitivity}
Secondary analyses examine patient groups using separate strata for age, Emergency Severity Index and recorded sex. The age bands are 18--44, 45--64 and 65--100 years. The acuity groups are Emergency Severity Index 1--2, 3 and 4--5. Missing acuity forms a separate group. Recorded sex categories, including Unknown, are retained as recorded. Each group uses its own marginal cut-offs, and we fit copula models in groups with at least 50 joint exceedances at the 95th-percentile cut-offs. Smaller groups, and unmapped or conflicting records, are described without copula inference. The age, acuity and sex analyses are performed separately, with pointwise intervals.

In the overall sample, sensitivity analyses for the six main families examine tie handling, repeated encounters and prior support. For tie handling, we average over eight independent random orderings within each marginal tie block. Fitted probabilities are averaged after transformation, and separately normalised posteriors are mixed with equal weights. Each bootstrap resample generates eight orderings and refits the models.

For repeated encounters, we refit the six families to all encounters using mid-ranks, with sandwich and patient-bootstrap intervals. The all-encounter joint-exceedance rate is also estimated at its own cut-offs and at the primary-sample cut-offs. Within each patient resample, we recalculate the cut-offs, empirical rates and copula fits. Sensitivity to prior support is examined by comparing the main and wider domains in Table~\ref{tab:support}. For both domains, we use the main-domain sandwich centre and covariance, so this comparison assesses the adjusted summaries under a common calibration.

A likelihood-based sensitivity analysis is conducted as well, by fitting the unrotated BB1 and BB7 copulas\citep{joe2014} by maximum pseudo-likelihood in every model-eligible primary group. Both families use 999 patient-bootstrap resamples and the same tail checks. Supplementary Appendix~\ref{app:families} gives their formulas and optimisation bounds, and Supplementary Appendix~\ref{app:numerics} describes the separate fitting procedures for BB1 and BB7.

\subsection{Simulation assessment}\label{sec:simulation}
In the first experiment, we assess interval performance in two simulation settings for each of the six main copula families. The first contains 512 independent observations with continuous standard normal margins. The second contains 256 independent patients with two observations each. With probability 0.4, the second pair equals the first pair for that patient. Otherwise, it is generated independently. Both margins are then rounded to the nearest 0.1. Each setting uses one specified parameter configuration, listed in Supplementary Appendix~\ref{app:numerics}.

We use 2,000 replicates per family and setting to compare unadjusted posterior intervals, sandwich-adjusted intervals, randomised-rank posterior mixtures and patient-bootstrap intervals for $R_{0.90}$ and $R_{0.95}$. Every bootstrap interval uses 999 patient resamples. Posterior summaries use 2,000 draws, allocated equally across eight orderings for a randomised-rank mixture. The bootstrap point estimate is the original-sample fitted probability. For the other methods, the point estimates are distribution medians.

Bias, interval width, coverage, computational failures and boundary fits are reported. Coverage is given both among available intervals and with failed intervals counted as noncoverage. For $B$ replicates and estimated coverage $\widehat c$, the Monte Carlo standard error is $\{\widehat c(1-\widehat c)/B\}^{1/2}$. At coverage 0.95 and $B=2,000$, it is approximately 0.0049. The generating and fitted copula families match in each setting.

A second experiment examined the choice of working family and the paired tail check. We generated data from the overall BB1 and BB7 fits to the primary sample and fitted the six main families to each replicate. We used 8,192 and 32,768 observations in each of two layouts, continuous independent pairs and rounded pairs with two visits per patient, and 107,387 rounded observations with one visit per patient. The last layout matches the primary sample size. Normal margins, rounding and within-patient copying followed the first experiment. Each setting used 2,000 replicates and 999 patient resamples per replicate. We recorded the pseudo-BIC winner, whether its fitted rate was closest to the known joint rate at the replicate's recorded cut-offs, and whether its paired 95\% interval for the fitted-minus-observed gap excluded zero. Separate Gaussian-independence controls assessed false flags with 512 to 32,768 observations. Supplementary Appendix~\ref{app:selection} defines the targets and gives the simulation parameters.

\subsection{Software}
The analyses use Python and pyvinecopulib version 1.0.1.\citep{vinecopulib}

\section{Simulation results}\label{sec:simulation-results}
\subsection{Coverage under the generating family}

We conducted a simulation study to assess the coverage of the interval methods for joint exceedance probabilities. Each observation comprised two simulated values, with standard normal marginal distributions and dependence specified by the generating copula. The first setting used independent continuous pairs, meaning that pairs were independent across observations. The second used rounded, repeated pairs, with two pairs per simulated patient and dependence between observations from the same patient.

The intervals were assessed against their stated (nominal) 95\% level by recording how often they contained the true probability that both variables exceeded their 90th or 95th marginal percentiles. The target was the probability under the generating copula, including when the measurements were rounded. Section~\ref{sec:simulation} gives the full simulation design, and Supplementary Appendix~\ref{app:numerics} lists the generating parameters for each family.

\begin{figure}[!htbp]
\centering
\includegraphics[width=\linewidth]{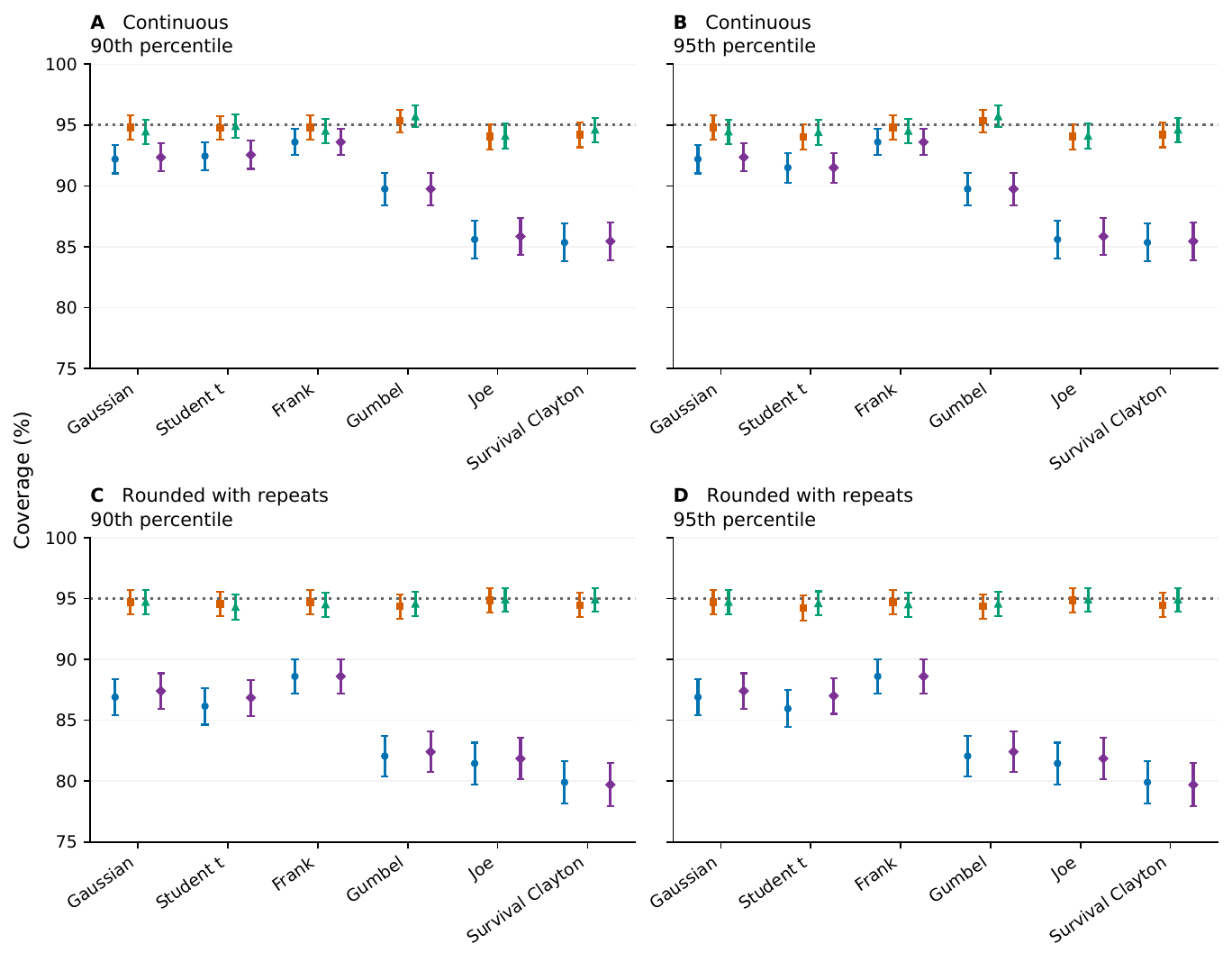}
\caption{Simulation coverage (\%) of nominal 95\% intervals. Circles denote uncorrected, squares sandwich-adjusted, triangles bootstrap and diamonds random-tie intervals.}\label{fig:simulation-coverage}
\par\noindent
\textbf{Alt text:} Four panels compare coverage across six copula families. The upper panels show independent continuous pairs and the lower panels rounded pairs with repeated observations. The left and right columns show the ninetieth and ninety-fifth percentiles. Sandwich-adjusted posterior and patient-cluster bootstrap coverage stays close to the dotted ninety-five percent line. Uncorrected and random-tie posterior coverage is lower, especially with rounded, repeated observations. Error bars show Monte Carlo uncertainty in coverage.
\end{figure}

Table~\ref{tab:simulation-coverage} and Figure~\ref{fig:simulation-coverage} give coverage over all 2,000 replicates per family and setting, with unavailable intervals counted as misses. The table gives coverage percentages for the joint-tail probability, with Monte Carlo standard errors in percentage points in parentheses. Panels A and B show continuous observations and Panels C and D show rounded, repeated observations, with the 90th percentile in Panels A and C and the 95th percentile in Panels B and D. The error bars in the figure show $\pm1.96$ Monte Carlo standard errors for the estimated coverage, and the dotted line marks 95\% coverage.

Across the six families and both percentiles, coverage ranged from 94.0\% to 95.7\% for the sandwich-adjusted posterior and the patient-cluster bootstrap. For the uncorrected and random-tie posteriors, it ranged from 79.7\% to 93.6\%. In every family, uncorrected intervals had lower coverage with rounded, repeated observations than with independent continuous observations. Randomising tied ranks did not restore nominal coverage.

Intervals from the adjusted posterior and patient bootstrap were wider on average than the uncorrected intervals in every setting. Only the Student $t$ fits gave unavailable intervals; all patient-bootstrap intervals were available.

Supplementary Appendix~\ref{app:simulation-details} (Tables~\ref{tab:sim-detail-c90}--\ref{tab:sim-detail-r95}) reports bias, root mean squared error, mean interval width and coverage among available intervals, together with the numbers of unavailable intervals and fits at a parameter bound for each family, setting and percentile.

\subsection{Model selection and joint rates under BB1 and BB7}
Within each replicate, pseudo-BIC selected the family once. The selected family was held fixed and its parameters were refitted in each of the 999 patient bootstrap samples. At 107,387 rounded observations, pseudo-BIC selected Student $t$ in 1,962 of 2,000 BB1 replicates and 1,997 of 2,000 BB7 replicates (Table~\ref{tab:selection}). At the 95th percentile, the selected family gave the closest of the six fitted rates in only seven BB1 replicates and one BB7 replicate. The paired interval excluded zero for the selected family in 83.6\% and 81.9\% of replicates, respectively (Monte Carlo standard errors 0.83 and 0.86 percentage points). At the 90th percentile, the selected family gave the closest rate much more often, in 57.85\% and 63.10\% of replicates.

\begin{table}[!htbp]
\caption{Model selection and the paired tail check at the 95th percentile.}\label{tab:selection}
\centering\begingroup\small
\setlength{\tabcolsep}{4pt}
\begin{tabular}{@{}llrrrrr@{}}\toprule
Generator & Layout & Observations & \shortstack{Mean joint\\count} & \shortstack{Student $t$\\selected (\%)} & \shortstack{Winner\\closest (\%)} & \shortstack{Winner\\flagged (\%)}\\ \midrule
BB1 & Continuous & 8,192 & 37.7 & 14.30 & 10.80 & 9.10\\
BB1 & Rounded, two visits & 8,192 & 32.2 & 15.30 & 11.85 & 5.00\\
BB1 & Continuous & 32,768 & 150.5 & 72.65 & 7.25 & 32.25\\
BB1 & Rounded, two visits & 32,768 & 133.1 & 66.30 & 6.95 & 21.30\\
BB1 & Rounded, one visit & 107,387 & 457.5 & 98.10 & 0.35 & 83.60\\
BB7 & Continuous & 8,192 & 37.0 & 19.85 & 12.20 & 8.30\\
BB7 & Rounded, two visits & 8,192 & 32.0 & 21.95 & 13.65 & 4.00\\
BB7 & Continuous & 32,768 & 148.6 & 84.95 & 6.50 & 28.60\\
BB7 & Rounded, two visits & 32,768 & 131.0 & 78.50 & 4.90 & 20.00\\
BB7 & Rounded, one visit & 107,387 & 449.4 & 99.85 & 0.05 & 81.90\\
\bottomrule\end{tabular}\endgroup
\par\smallskip\begin{minipage}{0.97\linewidth}\footnotesize Each row summarises 2,000 replicates. The winner has the lowest pseudo-BIC among six working families. Closest means the smallest absolute error in the fitted rate relative to the generating rate at the replicate's recorded cut-offs. A flag indicates that the paired 95\% bootstrap interval for the gap excludes zero. The continuous layout has one observation per patient. Monte Carlo standard errors for percentages are at most 1.12 percentage points.\end{minipage}
\end{table}

At the population 95th-percentile cut-offs for rounded data, the Student $t$ projections underestimated the generating rates by 11.3\% for BB1 and 11.2\% for BB7. Survival Clayton gave the closest population rate, with relative errors of $+1.0\%$ and $-1.2\%$. The population Student $t$ fits reached the upper limit of 30 degrees of freedom under both generators. At this sample size the Student $t$ fit reached the upper degrees-of-freedom limit in 1,731 BB1 and 1,422 BB7 replicates. The selected-family flag rates describe this selection and fitting procedure.

With 8,192 rounded observations from 4,096 patients, the selected family was flagged at the 95th percentile in 5.0\% of BB1 and 4.0\% of BB7 replicates. Both settings averaged about 32 joint events. In the Gaussian-independence controls at this percentile, the correctly specified Gaussian fit was flagged in 12.05\% of continuous samples and 23.30\% of rounded, repeated samples at 512 observations, with means of 1.22 and 0.98 joint events. Across the two layouts at 2,048 to 32,768 observations, false flags ranged from 3.85\% to 5.70\%. All pointwise intervals were available across the 36,000 replicates in the second experiment.

\section{Application to emergency department laboratory results}\label{sec:application}

This section examines how often creatinine and aspartate aminotransferase (AST) are both unusually high in the emergency department data. The observed joint rates are compared with those expected under independence, and joint probabilities are estimated using the rank-based Bayesian posterior. Model fit is assessed by comparing the observed rates with copula probabilities calculated at the maximum pseudo-likelihood estimates. Variation across patient groups and sensitivity to repeated encounters, tie handling and prior bounds are also examined.

\subsection{Joint exceedances in tested patients}
At their first qualifying encounter, 430 of the 107,387 patients with creatinine and AST measurements had both markers above their 95th-percentile cut-offs in the same specimen. These cut-offs were 1.70~mg/dL for creatinine and 90~U/L for AST, with 4.95\% and 4.97\% of patients above them, respectively. Both measurements exceeded their cut-offs in 4.00 per 1,000 patients, compared with 2.46 per 1,000 under independence. The observed rate was 1.63 times the rate expected under independence (95\% confidence interval 1.49 to 1.77). At the 90th percentile, 1,544 patients exceeded both cut-offs and the ratio was 1.45 (1.39 to 1.52).

Table~\ref{tab:app-empirical} gives the cut-offs and rates for the first-encounter and repeated-encounter samples. Both samples contain 107,387 patients, with rates expressed per 1,000 patients in the first sample and per 1,000 encounters in the second. Each rate under independence is the product of the observed proportions strictly above the two cut-offs. The pointwise 95\% confidence intervals in the table use 2,000 patient resamples.

\begin{table}[!htbp]
\caption{Creatinine and AST joint exceedances, with rates per 1,000.}\label{tab:app-empirical}
\centering\small
\setlength{\tabcolsep}{4pt}
\begin{tabular}{@{}rrrrrrr@{}}
\toprule
Percentile & \shortstack{Creatinine\\(mg/dL)} & \shortstack{AST\\(U/L)} & \shortstack{Joint\\count} & \shortstack{Observed rate\\(95\% CI)} & \shortstack{Rate under\\independence} & \shortstack{Ratio\\(95\% CI)}\\
\midrule
\multicolumn{7}{@{}l}{\textit{First qualifying encounter per patient (107,387 patients)}}\\
90 & 1.29 & 59 & 1,544 & 14.38 (13.59, 14.85) & 9.90 & 1.45 (1.39, 1.52)\\
95 & 1.70 & 90 & 430 & 4.00 (3.67, 4.37) & 2.46 & 1.63 (1.49, 1.77)\\
\addlinespace
\multicolumn{7}{@{}l}{\textit{All qualifying encounters (202,118 encounters)}}\\
90 & 1.49 & 63 & 2,479 & 12.27 (11.70, 12.84) & 9.79 & 1.25 (1.20, 1.31)\\
95 & 2.23 & 96 & 577 & 2.85 (2.56, 3.13) & 2.48 & 1.15 (1.03, 1.27)\\
\bottomrule
\end{tabular}
\end{table}

Supplementary Figure~\ref{fig:app-ratios} shows the corresponding comparisons within age, acuity and recorded sex groups, with separate cut-offs for each row. Bars show pointwise 95\% patient-bootstrap confidence intervals, and the vertical line marks independence. For patients aged 18--44 years, the 95th-percentile cut-offs were 1.19~mg/dL and 83~U/L. These cut-offs were 2.25~mg/dL and 88~U/L for those aged 65--100 years. The ratios were 1.84 (1.60 to 2.09) and 1.50 (1.28 to 1.76), respectively. The pointwise intervals for the ratios lay above one at both percentiles in every group with enough joint exceedances for modelling. Supplementary Appendix~\ref{app:application-details} (Table~\ref{tab:app-groups}) gives the counts and cut-offs, including those for groups with fewer than 50 joint exceedances at the 95th percentile.

\subsection{Agreement between fitted and observed joint rates}
Fitting the Gaussian copula across the full range of measurements gave a dependence parameter of 0.071 (sandwich-adjusted 95\% interval 0.064 to 0.077). Estimates of the probability that both measurements would exceed their nominal 95th percentiles differed between families. The sandwich-adjusted posterior median was 3.329 per 1,000 under the Gaussian copula (95\% interval 3.244 to 3.409) and 5.405 under the Student $t$ copula (5.240 to 5.563). These estimates refer to continuous percentiles, and the difference between the two medians was larger than either interval width. Appendix Table~\ref{tab:app-prior} gives the estimates for all six families.

\begin{figure}[!htbp]
\centering
\includegraphics[width=.9\linewidth]{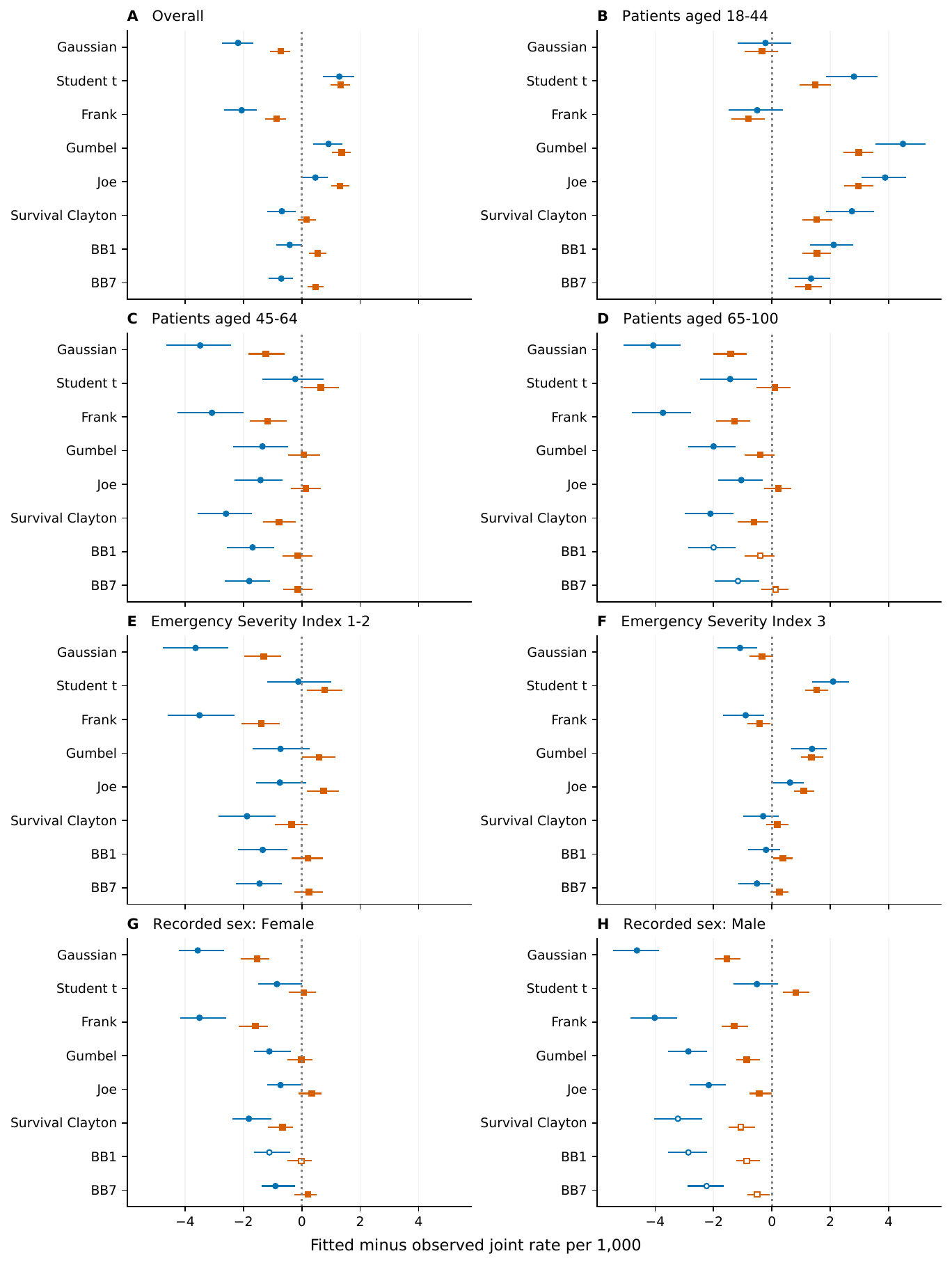}
\caption{Fitted minus observed joint rates per 1,000 patients. Circles and squares denote the 90th and 95th percentiles, and open markers indicate parameter bounds.}\label{fig:app-gaps}
\par\noindent
\textbf{Alt text:} Eight panels compare the families overall and within patient groups. Overall, the Gaussian and Frank copulas underestimate the rate at both cut-offs. The Student $t$ and Gumbel copulas overestimate it at both.
\end{figure}

To check these fits against the recorded measurements, we evaluated each family at the observed marginal cut-offs (Figure~\ref{fig:app-gaps}). Panel A shows results based on the first qualifying encounter for each patient. Panels B--H show the patient groups with enough joint exceedances for modelling: ages 18--44, 45--64 and 65--100 years (B--D), Emergency Severity Index 1--2 and 3 (E and F), and patients recorded as female and male (G and H). Each panel includes the six main copula families and BB1 and BB7. Points show fitted minus observed joint exceedances per 1,000 patients, so positive differences indicate overestimation. Bars show pointwise 95\% confidence intervals from paired patient resamples, and open markers identify the six fits that reached a parameter bound: BB1 and BB7 for ages 65--100, BB1 for patients recorded as female, and survival Clayton, BB1 and BB7 for patients recorded as male.

At the observed 95th-percentile cut-offs, the joint rate was 4.00 per 1,000 patients. The Gaussian fit differed from the observed rate by $-0.726$ per 1,000 patients (95\% confidence interval $-1.084$ to $-0.409$), or $-18.1\%$ relative to the observed rate. The Student $t$ copula had the lowest pseudo-BIC overall and survival Clayton the highest (Table~\ref{tab:app-model-checks}). The Student $t$ fit exceeded the observed rate by 1.335 per 1,000 patients (95\% confidence interval 0.976 to 1.642), or 33.3\% relative to the observed rate. The survival Clayton fit exceeded the observed rate by 0.168 per 1,000 patients (95\% confidence interval $-0.140$ to 0.477), or 4.2\% relative to the observed rate. At the 90th-percentile cut-offs, the survival Clayton fit differed from the observed rate by $-0.681$ per 1,000 patients (95\% confidence interval $-1.191$ to $-0.208$), or $-4.7\%$ relative to the observed rate. Relative errors are point estimates of $100(\mathrm{fitted}-\mathrm{observed})/\mathrm{observed}$. In the parameterisations of Supplementary Appendix~\ref{app:families}, the overall maximum pseudo-likelihood estimates were $\widehat\rho=0.071$ for the Gaussian copula, $(\widehat\rho,\widehat\nu)=(0.087,9.09)$ for the Student $t$ copula, $\widehat\theta=0.570$ for Frank, 1.049 for Gumbel, 1.053 for Joe and 0.071 for survival Clayton, and $(\widehat\theta,\widehat\delta)=(0.047,1.032)$ for BB1 and $(1.034,0.066)$ for BB7.

Across the six main families, the pointwise intervals for fitted minus observed excluded zero in 71 of 96 comparisons, and in 69 of 94 comparisons across the 47 fits not at a parameter bound (Table~\ref{tab:app-model-checks}). Only four of the 48 family and group combinations in that table had intervals containing zero at both percentiles. These were Gaussian for ages 18--44, Gumbel for Emergency Severity Index 1--2, survival Clayton for Emergency Severity Index 3 and Student $t$ for patients recorded as female. Every BB1 and BB7 fit not at a parameter bound missed at least one observed joint rate. For patients recorded as male, the survival Clayton estimate reached the lower parameter bound and its sandwich-adjusted interval was unavailable.

\subsection{Repeated encounters, ties and prior bounds}
To assess sensitivity to repeat visits, we included all 202,118 qualifying encounters with equal weight. The 95th-percentile cut-offs rose to 2.23~mg/dL for creatinine and 96~U/L for AST. The observed rate was 2.85 per 1,000 encounters, or 1.15 times the independence rate (95\% confidence interval 1.03 to 1.27). Using the first-encounter cut-offs for all encounters gave a ratio of 1.26 (1.17 to 1.34). Thus the ratio was smaller when repeat encounters were included even when the cut-offs were held fixed.

We also examined sensitivity to tie handling and prior bounds. Random ordering of tied values and mid-ranks gave similar fitted rates. At the 95th percentile, the Student $t$ estimates were 5.33 and 5.34 per 1,000, respectively. Widening the prior bounds changed the sandwich-adjusted posterior medians by less than 0.01 per 1,000 for every family at both nominal percentiles. To compare the adjusted summaries, we use the same sandwich centre and covariance from the main domain for both prior supports. The unadjusted posterior medians also changed by less than 0.01 per 1,000. Supplementary Appendix~\ref{app:application-details} gives the fitted rates under the alternative tie and encounter rules (Table~\ref{tab:app-ties}), the posterior estimates under both prior supports (Table~\ref{tab:app-prior}) and the paired comparison of the encounter samples (Table~\ref{tab:app-visits}).

\section{Discussion}\label{sec:discussion}

At the observed 95th-percentile cut-offs, where the observed joint rate was 4.00 per 1,000 patients, the Student $t$ copula had the lowest pseudo-BIC overall but overestimated the rate by 1.335 per 1,000 (33.3\%). The survival Clayton copula, with the highest pseudo-BIC, overestimated the same rate by 0.168 per 1,000 (4.2\%), although it underestimated the 90th-percentile rate by 0.681 per 1,000 (95\% confidence interval $-1.191$ to $-0.208$), and the Gaussian copula underestimated the 95th-percentile rate by 0.726 per 1,000 (18.1\%). These comparisons sit alongside two broader findings. Creatinine and AST were high together more often than expected under independence in the emergency department data, and in the simulations, sandwich-adjusted posterior and patient-cluster bootstrap intervals contained the true joint probability more often than unadjusted posterior intervals, with coverage close to the intended level.

The second simulation helps interpret the difference between overall fit and agreement at a particular cut-off. At the primary sample size, Student $t$ usually ranked first among the six working families, yet survival Clayton gave the closest population rate at the 95th percentile under both generators. The Student $t$ error changed direction between the simulated and recorded data. Thus, the model ranking alone gave little guidance about either the size or the direction of the error at this threshold. The two generators were fitted to the same cohort and deliberately excluded from the candidate set. These results describe selection among misspecified families. They also show why absence of a flag is difficult to interpret when joint events are sparse. The independence controls assess a fixed Gaussian model only, and the size comparisons do not establish a minimum event count for subgroup analysis.

At the first qualifying encounter, 4 per 1,000 patients had both measurements above their 95th-percentile cut-offs, compared with 2.46 per 1,000 under independence. The ratio was 1.63 (95\% confidence interval 1.49 to 1.77). Within the age, acuity and recorded sex groups with enough joint exceedances for modelling, the 95th-percentile ratios ranged from 1.43 to 1.84 (Supplementary Figure~\ref{fig:app-ratios}).

Including repeat encounters lowered the ratio from 1.63 to 1.26 at the first-encounter cut-offs and to 1.15 when the cut-offs were re-estimated. Most of the decrease therefore occurred at fixed cut-offs. The all-encounter estimate gives greater weight to frequent visitors. Supplementary Appendix~\ref{app:application-details} discusses this difference.

Earlier reports describe rhabdomyolysis, hypoxic liver disease and sepsis,\citep{weibrecht2010,tapper2015,drolz2016,singer2016} while the studies relating AST and creatinine after liver transplantation or myocardial infarction examined selected patient cohorts.\citep{leithead2013,park2016} Our analysis instead describes joint elevations in the tested emergency department population; co-occurrence does not identify its cause. Independence provides a reference for measuring co-occurrence, without an assumption of biological independence. The decile comparison in Figure~\ref{fig:data-deciles} also shows patterns beyond the upper tail. The lowest creatinine group combined with the highest AST group had a ratio of 1.25, which could reflect conditions such as cirrhosis, in which reduced muscle mass lowers serum creatinine.\citep{simonetto2020} The highest creatinine group was close to independence with the lowest AST group (1.03) and below it with AST groups 2--4 (0.80 to 0.84). The excess where both measurements were in their lowest groups (1.40) is consistent with co-occurrence in both tails. The Student $t$ copula, whose dependence is the same in both tails, ranked first on pseudo-BIC but overestimated the upper-tail rate.

The tested population is close to the adult emergency department population that has blood drawn. In our department the triage nurse sends a comprehensive metabolic panel for most adults at arrival, whatever the presenting complaint, and basic metabolic panels are rarely used, so creatinine and AST come from the same panel and AST is measured without anyone having to suspect liver injury. Of the 285,024 initially eligible adult encounters, 202,413 (71.0\%) had a structurally qualifying same-specimen creatinine and AST pair before the numeric exclusions. Because the decision to test is made at triage, before a diagnosis and largely apart from the clinical question, AST testing largely does not depend on suspicion of liver injury. We did not assess how selection into blood testing affects the observed dependence. Pooling patients of very different acuity can still produce some dependence between two analytes, and the acuity-stratified ratios of 1.78 for Emergency Severity Index 1--2 and 1.43 for Index 3 address this only in part.

The observed joint rates can be estimated directly from the measurements and provide benchmarks for assessing the fitted copulas (Figure~\ref{fig:app-gaps}; Table~\ref{tab:app-model-checks}). The discrepancies described above show why interval calibration must be accompanied by checks that fitted models reproduce the rates of interest. The Gaussian and Student $t$ copulas also differ in their limiting upper-tail dependence for correlations strictly between $-1$ and $1$,\citep{demarta2005} but our analysis assesses agreement at the observed cut-offs. Every BB1 and BB7 fit not at a parameter bound also missed at least one observed joint rate. For each fitted-minus-observed difference, we calculated a pointwise interval by refitting the copula and re-estimating the observed rate in the same patient bootstrap resamples, with each interval applying to one family, group and threshold. Among the six main families, these intervals excluded zero in 71 of 96 comparisons, and in 69 of 94 comparisons across the 47 fits not at a parameter bound. Nominal coverage has not been established for the bootstrap intervals at parameter bounds, so comparisons for the six fits at a bound (one survival Clayton, five BB1 or BB7) remain descriptive.

Because the true joint probability was known in the simulations, we could check how often each interval contained it. Across the settings examined, coverage of intervals with a stated 95\% level ranged from 94.0\% to 95.7\% for the sandwich-adjusted posterior and the patient-cluster bootstrap. It ranged from 79.7\% to 93.6\% for the uncorrected and random-tie posteriors (Table~\ref{tab:simulation-coverage}). The sandwich adjustment and patient bootstrap both account for uncertainty in the ranks and for repeated measurements within patients. Each simulation in the first experiment fitted the generating family and used one parameter setting per family, so the coverage findings apply to the settings studied. Under the assumptions of Proposition~\ref{prop:calibration}, the sandwich adjustment calibrates uncertainty around the working-model target $R(\boldsymbol z_\ast)$. It does not remove bias from rounding or an incorrect copula family. Coverage of the latent probability in the rounded simulations is an empirical result for the settings studied.

The clinical interpretation remains limited to patients who had both tests on the same specimen at one emergency department. The cut-offs describe high values in this tested population and are not clinical definitions of injury. We did not assess whether patients with both elevations had kidney or liver injury. The laboratory extract used here had no result-comment field, so specimens affected by haemolysis could not be identified. The joint rate was not evaluated as a tool for diagnosis, outcome prediction or treatment decisions. The group comparisons use separate strata and group-specific cut-offs, so they do not estimate the joint effects of age, acuity and recorded sex.

Rounding can place several measurements exactly at a percentile cut-off. Only measurements strictly above that cut-off are counted, so ties can leave fewer than 5\% of measurements above the 95th percentile. Ties can also affect rank-based estimates of copula dependence.\citep{genestneslehova2007,li2020} Agreement between mid-ranks and random tie-breaking does not establish the absence of tie bias.\citep{li2020} A continuous copula approximates the dependence between these rounded measurements. In a continuous model, exactly 5\% of the marginal distribution lies above its 95th percentile. The joint probability calculated at these model percentiles can therefore differ from the empirical joint rate at the recorded cut-offs. Our checks compare fitted and observed rates at the recorded cut-offs so that both refer to the same exceedance event. Cut-offs are re-estimated in each resample, and bootstrap behaviour can be irregular when the percentile level equals an endpoint of a jump in the population distribution function. In the first-encounter sample, at the 90th percentile, each laboratory cut-off took four distinct values across the 2,000 resamples (creatinine 1.28 to 1.31 mg/dL; AST 58 to 61 U/L), and movement between tied values can contribute to the asymmetry of some 90th-percentile intervals. The first simulation assessed $R_q$ intervals. The second describes the paired gap intervals at the recorded cut-offs, including after family selection and at parameter bounds. Neither experiment assessed coverage of the empirical ratio intervals.

Future work should repeat the comparison in emergency departments with different testing practices and examine other forms of misspecification, including stronger dependence and different patterns of rounding. Further clinical studies could also assess whether joint elevations help predict patient outcomes in this tested population. The same question extends to more than two measurements, for example the creatinine, bilirubin, lactate and platelet changes that accompany sepsis-associated organ dysfunction, which would call for vine or factor copulas rather than a bivariate model.

\clearpage
\bibliographystyle{unsrtnat}
\bibliography{references}
\clearpage
\section*{Statements and Declarations}
\subsection*{Ethical considerations}
The Stanford University Institutional Review Board approved this retrospective study of existing electronic health records (protocol 69107).
\subsection*{Consent to participate}
The Stanford University Institutional Review Board waived informed consent for the retrospective study.
\subsection*{Consent for publication}
Not applicable. The article reports aggregate results.
\subsection*{AI-use statement}
The authors used ChatGPT 6 Astra (OpenAI) and Claude Opus 5.5 (Anthropic) for drafting and editing the manuscript. The authors reviewed all output and take full responsibility for the work.
\subsection*{Funding}
This research received no specific grant from any funding agency.
\subsection*{Declaration of conflicting interest}
The authors declared no potential conflicts of interest with respect to the research, authorship or publication of this article.
\subsection*{Data and code availability}
Patient-level data cannot be shared under institutional data-use agreements. Analysis code is available from the corresponding author on request.
\end{document}


{\noindent\sffamily\bfseries\Large Supplementary material\par}
\medskip\noindent Sandwich-adjusted Bayesian copula inference and model checks for joint laboratory extremes in emergency care\par\medskip
\noindent Appendices A--C give the model definitions, calibration argument and numerical methods. Appendix D gives the matched-family simulation results. Appendices E and F give the additional application results and record-selection rules. Appendix G gives the design and results for model selection and the paired tail check.
\appendix
\renewcommand{\thefigure}{S\arabic{figure}}
\renewcommand{\thetable}{S\arabic{table}}
\renewcommand{\theequation}{S\arabic{equation}}
\renewcommand{\theproposition}{S\arabic{proposition}}
\section{Copula families and parameter domains}\label{app:families}

This appendix specifies the copula distribution functions and parameter ranges used in the analyses. It also gives the BB1 log-density used for stable evaluation near its Gumbel limit.

\begin{figure}[!htbp]
\centering
\begin{tikzpicture}[
  box/.style={draw=black!70, rounded corners=2pt, align=center, font=\small, text width=46mm, minimum height=11mm, inner sep=3pt},
  key/.style={box, draw=flowblue, line width=0.9pt},
  arr/.style={-{Stealth[length=2mm]}, black!70}]
\node[box] (data) at (0,0) {Same-specimen creatinine and AST (Section~\ref{sec:data})};
\node[box] (ranks) at (-3.4,-1.8) {Mid-rank pseudo-observations};
\node[box] (obs) at (3.4,-1.8) {Observed joint rates at the recorded cut-offs (Section~\ref{sec:empirical})};
\node[box] (post) at (-3.4,-3.6) {Six main families: pseudo-likelihood and restricted Jeffreys prior give the unadjusted posterior};
\node[box] (mpl) at (3.4,-3.6) {Maximum pseudo-likelihood fit evaluated at the same cut-offs};
\node[key] (sand) at (-3.4,-5.4) {Sandwich adjustment for estimated margins and patient clustering (Section~\ref{sec:calibration})};
\node[key] (gap) at (3.4,-5.4) {Fitted minus observed rates, with paired patient-resample intervals (Section~\ref{sec:adequacy})};
\node[box] (rq) at (-3.4,-7.2) {Calibrated intervals for $R_q$};
\node[box] (boot) at (3.4,-7.2) {Patient-cluster bootstrap intervals for $R_q$};
\draw[arr] (data.south) -- ++(0,-0.4) -| (ranks.north);
\draw[arr] (data.south) -- ++(0,-0.4) -| (obs.north);
\draw[arr] (ranks) -- (post);
\draw[arr] (ranks.east) -- (mpl.north west);
\draw[arr] (obs) -- (mpl);
\draw[arr] (post) -- (sand);
\draw[arr] (sand) -- (rq);
\draw[arr] (mpl) -- (gap);
\draw[arr] (data.east) -- (6.4,0) -- node[sloped, above, font=\footnotesize]{resample patients, rebuild ranks, refit} (6.4,-7.2) -- (boot.east);
\draw[arr, dashed] (boot) -- node[above, font=\footnotesize]{comparison} (rq);
\end{tikzpicture}
\caption{Estimation, interval calibration and model checks.}\label{fig:methods-flow}
\par\smallskip
\begin{minipage}{0.95\linewidth}\footnotesize
\textbf{Alt text:} Flow diagram. Same-specimen measurements lead to mid-rank pseudo-observations and to observed joint rates. On the left, the pseudo-observations give an unadjusted posterior for the six main families, which a sandwich adjustment turns into calibrated intervals. On the right, a maximum pseudo-likelihood fit evaluated at the observed cut-offs is compared with the observed rates using paired patient resamples. Resampling patients, rebuilding ranks and refitting gives bootstrap intervals that are compared with the calibrated intervals.
\end{minipage}
\end{figure}

\subsection{Distribution functions}
We use the pyvinecopulib parameterisations.\citep{vinecopulib} The formulas below apply to $0<u,v<1$, with boundary values defined by continuity. The same parameter symbol can have a different interpretation in different families.

For the Gaussian and Student $t$ families,
\begin{align}
C_{\rho}^{\mathrm G}(u,v)&=\Phi_{\rho}\{\Phi^{-1}(u),\Phi^{-1}(v)\},\label{eq:gaussian}\\
C_{\rho,\nu}^{t}(u,v)&=T_{\rho,\nu}\{t_\nu^{-1}(u),t_\nu^{-1}(v)\}.\label{eq:student}
\end{align}
Here $\Phi$ is the standard normal distribution function, and $\Phi_\rho$ is the bivariate normal distribution function with correlation $\rho$. The functions $t_\nu$ and $T_{\rho,\nu}$ are the univariate and bivariate Student $t$ distribution functions. Both use $\nu$ degrees of freedom, and the bivariate function has dependence parameter $\rho$. In both families, $\rho$ controls dependence in the bivariate distribution that defines the copula.

The Frank, Gumbel and Joe distribution functions are
\begin{align}
C_\theta^{\mathrm F}(u,v)
&=-\theta^{-1}\log\left[1+
\frac{\{\exp(-\theta u)-1\}\{\exp(-\theta v)-1\}}{\exp(-\theta)-1}\right],\label{eq:frank}\\
C_\theta^{\mathrm{Gu}}(u,v)
&=\exp\left[-\{(-\log u)^\theta+(-\log v)^\theta\}^{1/\theta}\right],\label{eq:gumbel}\\
C_\theta^{\mathrm J}(u,v)
&=1-\{(1-u)^\theta+(1-v)^\theta-(1-u)^\theta(1-v)^\theta\}^{1/\theta}.\label{eq:joe}
\end{align}
The Frank formula applies when $\theta\ne0$, and its continuous limit at zero is $uv$. Gumbel and Joe require $\theta\ge1$.

The survival Clayton family is obtained by rotating the Clayton family through 180 degrees. Its distribution function is
\begin{equation}\label{eq:survivalclayton}
C_\theta^{\mathrm{SC}}(u,v)
=u+v-1+\{(1-u)^{-\theta}+(1-v)^{-\theta}-1\}^{-1/\theta},\qquad \theta>0.
\end{equation}

The BB1 and BB7 families are given by Joe (Sections 4.17.1 and 4.23.1).\citep{joe2014} The BB1 distribution function is
\begin{equation}\label{eq:bb1}
C_{\theta,\delta}^{\mathrm{BB1}}(u,v)
=\left[1+\{(u^{-\theta}-1)^\delta+(v^{-\theta}-1)^\delta\}^{1/\delta}\right]^{-1/\theta},
\qquad \theta>0,\quad\delta\ge1.
\end{equation}
As $\theta\downarrow0$, the BB1 distribution function converges to the Gumbel copula with parameter $\delta$. We include this limit in the fitted BB1 domain.

For BB7, define $A_u=1-(1-u)^\theta$ and $A_v=1-(1-v)^\theta$. The distribution function is
\begin{equation}\label{eq:bb7}
C_{\theta,\delta}^{\mathrm{BB7}}(u,v)
=1-\left[1-\{A_u^{-\delta}+A_v^{-\delta}-1\}^{-1/\delta}\right]^{1/\theta},
\qquad \theta\ge1,\quad\delta>0.
\end{equation}
For fitting, we restrict the BB7 domain to a lower bound of $\delta=0.01$.

\Needspace{14\baselineskip}
\subsection{A stable BB1 density}
To evaluate the BB1 density near $\theta=0$, write $x=-\log u$, $y=-\log v$ and
\[
A=\frac{\exp(\theta x)-1}{\theta},\qquad
B=\frac{\exp(\theta y)-1}{\theta},\qquad
r=(A^\delta+B^\delta)^{1/\delta},\qquad L=\log(1+\theta r).
\]
\Needspace{9\baselineskip}
At $\theta=0$, take $A=x$, $B=y$ and $\log C=-r$ by continuity. Otherwise $\log C=-L/\theta$. Differentiating equation~\eqref{eq:bb1} gives
\begin{equation}\label{eq:bb1logdensity}
\begin{split}
\log c^{\mathrm{BB1}}_{\theta,\delta}(u,v)
={}&\log C-2L+(2-2\delta)\log r
+(\delta-1)(\log A+\log B)\\
&+(\theta+1)(x+y)
+\log\{1+\theta\delta+(\delta-1)/r\}.
\end{split}
\end{equation}
We evaluate $\exp(a)-1$ with \texttt{expm1} and $\log(1+a)$ with \texttt{log1p}, which provide numerically stable forms of these expressions.

\subsection{Parameter restrictions}
\begin{table}[!htbp]
\caption{Main and wider parameter ranges for posterior inference.}\label{tab:support}
\centering
\begin{tabular}{llll}
\toprule
Family & Parameter & Main support & Wider support\\
\midrule
Gaussian & $\rho$ & $[-0.95,0.95]$ & $[-0.98,0.98]$\\
Student $t$ & $\rho$ & $[-0.95,0.95]$ & $[-0.98,0.98]$\\
Student $t$ & $\nu$ & $[2.01,30]$ & $[2.001,49.9]$\\
Frank & $\theta$ & $[-15,15]$ & $[-25,25]$\\
Gumbel & $\theta$ & $[1.001,8]$ & $[1.0001,12]$\\
Joe & $\theta$ & $[1.001,8]$ & $[1.0001,12]$\\
Survival Clayton & $\theta$ & $[0.001,8]$ & $[0.0001,12]$\\
\bottomrule
\end{tabular}
\end{table}

Table~\ref{tab:support} gives the main and wider parameter ranges for the six Bayesian models. The restriction is applied jointly to both Student $t$ parameters. For the likelihood-based sensitivity analyses, the optimisation bounds are $\theta\in[0,7]$, $\delta\in[1,7]$ for BB1 and $\theta\in[1,6]$, $\delta\in[0.01,25]$ for BB7.

\section{Rank correction and local calibration}\label{app:calibration}

This appendix gives the empirical-margin score correction and the adjustment of posterior draws. It then states the assumptions under which the adjusted intervals have asymptotic nominal coverage for a copula tail functional.

\subsection{Calibration target}
For independent observations with continuous margins, this correction has the first-order form of Genest et al.,\citep{genest1995} and rank-based estimation of this kind requires no knowledge of the marginal distributions.\citep{genest1995,tsukahara2005} We use mid-ranks for ties and sum the corrected scores within patients. To preserve the parameter support, we centre and rescale posterior draws in logit coordinates of the bounded parameter domain. The resulting sandwich-adjusted quasi-posterior matches the estimated sampling covariance locally. Proposition~\ref{prop:calibration} in Supplementary Appendix~\ref{app:calibration} gives conditions under which the sandwich-adjusted equal-tailed intervals have asymptotic nominal coverage for the working copula tail probability. This working target is $R(\boldsymbol z_\ast)$, where $\boldsymbol z_\ast$ maximises the expected log copula density evaluated at the population mid-distribution transforms over the specified parameter domain, using the encounter-weighted distribution in the repeated-visit analysis. With continuous margins and a misspecified copula family, $\boldsymbol z_\ast$ is the pseudo-true parameter value, which minimises the Kullback--Leibler divergence from the true copula density.\citep{chen2006} Under the interior regularity conditions of Proposition~\ref{prop:calibration}, the adjusted equal-tailed interval for $R_q$ is first-order equivalent to a patient-cluster sandwich Wald interval. The adjusted draws give quantiles of the nonlinear tail probability directly, while respecting the parameter bounds.

\subsection{Empirical-margin contribution to the score}
To account for the effect of each observation on the estimated margins, define $h(a,b)=\mathbf 1(a<b)+\tfrac12\mathbf 1(a=b)$. The mid-rank pseudo-observations can be written exactly as
\begin{equation}\label{eq:midrank}
\widehat U_i=\frac{1/2+\sum_{l=1}^N h(X_l,X_i)}{N+1},\qquad
\widehat V_i=\frac{1/2+\sum_{l=1}^N h(Y_l,Y_i)}{N+1}.
\end{equation}
Let $\boldsymbol s_i$ be the derivative of $\log c_{\boldsymbol\theta(\boldsymbol z)}(\widehat U_i,\widehat V_i)$ with respect to $\boldsymbol z$, evaluated at $\widehat{\boldsymbol z}$. The vectors $\boldsymbol d_{i1}$ and $\boldsymbol d_{i2}$ are the derivatives of this score with respect to the first and second components of the pseudo-observation pair, respectively. We use these derivatives to account for uncertainty in the estimated margins through the correction
\begin{equation}\label{eq:influence}
\begin{split}
\boldsymbol a_l&=\boldsymbol s_l+
\frac{1}{N+1}\sum_{i=1}^N
\{\boldsymbol d_{i1}h(X_l,X_i)+\boldsymbol d_{i2}h(Y_l,Y_i)\},\\
\widehat{\boldsymbol\psi}_l&=\boldsymbol a_l-N^{-1}\sum_{j=1}^N\boldsymbol a_j.
\end{split}
\end{equation}
Within each margin, observation $l$ contributes fully to the ranks of observations with larger values and contributes half as much to the ranks of observations tied with it. Differentiating equation~\eqref{eq:midrank} gives the two margin terms in equation~\eqref{eq:influence}, and centring removes terms common to all observations. We compute the correction using sorted values and cumulative sums, treating observations with the same value as a tie block. We then sum the corrected contributions within patients to obtain the covariance in equation~\eqref{eq:sandwich} for the mid-rank estimator.

\subsection{Transformation of posterior draws}
To keep the adjusted draws within the parameter bounds, we centre and rescale them on a logit scale before transforming them back. Let $(l_j,u_j)$ be the bounds of coordinate $z_j$. Define
\begin{equation}\label{eq:logit}
g_j(z_j)=\log\left(\frac{z_j-l_j}{u_j-z_j}\right),\qquad
\boldsymbol\eta=g(\boldsymbol z),\qquad
D=\left.\frac{\partial g}{\partial\boldsymbol z^{\mathsf T}}\right|_{\widehat{\boldsymbol z}}.
\end{equation}
The matrix $D$ is diagonal, with entries
$(u_j-l_j)/\{(\widehat z_j-l_j)(u_j-\widehat z_j)\}$.
For the unadjusted posterior draws $\boldsymbol z^{(b)}$, calculate the mean $\overline{\boldsymbol\eta}$ and covariance $S_\eta$ of $g(\boldsymbol z^{(b)})$. For a positive-definite matrix $Q$, let $L(Q)$ denote its lower Cholesky factor. We adjust the draws by
\begin{equation}\label{eq:adjustment}
\begin{split}
A&=L(D\widehat V_zD^{\mathsf T})L(S_\eta)^{-1},\\
\boldsymbol\eta^{(b)}_{\mathrm{adj}}
&=g(\widehat{\boldsymbol z})+A\{\boldsymbol\eta^{(b)}-\overline{\boldsymbol\eta}\},\\
\boldsymbol z^{(b)}_{\mathrm{adj}}&=g^{-1}(\boldsymbol\eta^{(b)}_{\mathrm{adj}}).
\end{split}
\end{equation}
This construction gives the adjusted draws sample covariance $D\widehat V_zD^{\mathsf T}$ on the $\boldsymbol\eta$ scale. Inverting $g$ returns the draws to the $\boldsymbol z$ scale while preserving the parameter bounds. The covariance matching on this scale is local. This implements the covariance adjustment in bounded coordinates.\citep{shaby2014}

\subsection{Scope of the large-sample argument}
For rounded data, define the population mid-distributions
$F_X^{\mathrm m}(x)=\Pr(X<x)+\tfrac12\Pr(X=x)$ and $F_Y^{\mathrm m}$ analogously. The population target is the parameter that maximises the expected log copula density at these transforms. In the repeated-visit analysis, the expectation uses the encounter-weighted distribution. Denote the transformed maximiser by $\boldsymbol z_\ast$. Proposition~\ref{prop:calibration} gives conditions for local calibration at this target.

\begin{proposition}[Local calibration of a tail functional]\label{prop:calibration}
Suppose the number of independent patient clusters $M$ tends to infinity and the following conditions hold.
\begin{enumerate}
\item The population maximiser $\boldsymbol z_\ast$ is unique and lies in the interior of a fixed bounded domain. The estimator $\widehat{\boldsymbol z}$ is consistent. Also, $N/M$ converges to a finite positive constant.
\item The corrected estimating equation has the expansion
\[
\sqrt M(\widehat{\boldsymbol z}-\boldsymbol z_\ast)
=H^{-1}M^{-1/2}\sum_{m=1}^M\boldsymbol\Psi_m+o_p(1),
\]
where $M^{-1/2}\sum_m\boldsymbol\Psi_m\Rightarrow\mathcal N(\boldsymbol0,J)$ and $H_N/M\to_p H$. The matrices $H$ and $J$ are positive definite. The covariance estimator satisfies $M\widehat V_z\to_p\Sigma=H^{-1}JH^{-\mathsf T}$.
\item The log pseudo-likelihood has a locally uniform quadratic expansion at $\widehat{\boldsymbol z}$ with curvature $H_N$. The prior density is continuous and positive at $\boldsymbol z_\ast$. The unadjusted posterior concentrates in that neighbourhood. Its centred, $\sqrt M$-scaled distribution converges to $\mathcal N(\boldsymbol0,H^{-1})$, with convergence of its first two moments. Under this posterior, the first two moments of $\sqrt M\{g(\boldsymbol z)-g(\widehat{\boldsymbol z})\}$ converge in probability to the corresponding moments of $\mathcal N(\boldsymbol0,D_\ast H^{-1}D_\ast^{\mathsf T})$, where $D_\ast$ is the Jacobian of $g$ at $\boldsymbol z_\ast$. Numerical integration and simulation errors are negligible at this scale.
\item $R(\boldsymbol z)=R_q\{\boldsymbol\theta(\boldsymbol z)\}$ is continuously differentiable near $\boldsymbol z_\ast$, with $\nabla R(\boldsymbol z_\ast)^{\mathsf T}\Sigma\nabla R(\boldsymbol z_\ast)>0$.
\end{enumerate}
Then the sampling limit of $\sqrt M\{R(\widehat{\boldsymbol z})-R(\boldsymbol z_\ast)\}$ is centred normal with variance
\begin{equation}\label{eq:delta}
\nabla R(\boldsymbol z_\ast)^{\mathsf T}\Sigma\nabla R(\boldsymbol z_\ast).
\end{equation}
Conditionally on the data, the adjusted distribution of $\sqrt M\{R(\boldsymbol z_{\mathrm{adj}})-R(\widehat{\boldsymbol z})\}$ has the same limit. Its equal-tailed intervals consequently have asymptotic nominal coverage for $R(\boldsymbol z_\ast)$, the tail functional at the population target.
\end{proposition}

\begin{proof}
Condition (b) and the delta method give the sampling limit. Write
$K_\ast=D_\ast H^{-1}D_\ast^{\mathsf T}$ and
$T_\ast=D_\ast\Sigma D_\ast^{\mathsf T}$.
By condition (c) and the delta method, the conditional distribution of
$\sqrt M\{g(\boldsymbol z)-g(\widehat{\boldsymbol z})\}$ converges to
$\mathcal N(\boldsymbol0,K_\ast)$. The moment and numerical error
assumptions in condition (c) give
$\sqrt M\{\overline{\boldsymbol\eta}-g(\widehat{\boldsymbol z})\}\to_p\boldsymbol0$
and $M S_\eta\to_p K_\ast$.

Since $D\to_p D_\ast$ and $M\widehat V_z\to_p\Sigma$, the adjustment
matrix $A$ in equation~\eqref{eq:adjustment} converges in probability to
$L(T_\ast)L(K_\ast)^{-1}$. The centred, $\sqrt M$-scaled adjusted draws
therefore have conditional limit $\mathcal N(\boldsymbol0,T_\ast)$ on
the logit scale. The derivative of $g^{-1}$ at $g(\boldsymbol z_\ast)$
is $D_\ast^{-1}$, giving the conditional limit
$\mathcal N(\boldsymbol0,\Sigma)$ on the $\boldsymbol z$ scale.
Applying the delta method to $R$ gives the variance in
equation~\eqref{eq:delta}. The sampling and conditional limits are the
same normal distribution with positive variance, so the equal-tailed
intervals have asymptotic nominal coverage.
\end{proof}

Proposition~\ref{prop:calibration} requires an interior population maximiser, positive-definite information and the stated score and posterior limits. Verifying these conditions requires control of score derivatives near the edges of the unit square, cluster sizes and the marginal distribution. The coverage conclusion applies to $R(\boldsymbol z_\ast)$. With fixed rounding or model misspecification, this target can differ from a latent copula tail probability or the observed joint-exceedance rate.

The proposition does not establish bootstrap coverage for empirical quantile-based probabilities, which requires separate regularity conditions. A sample quantile can behave irregularly when $q$ equals the endpoint of a jump in the population distribution function. Boundary fits also require care. Six fits in Figure~\ref{fig:app-gaps} reached a parameter bound, and for those fits we have not established nominal 95\% coverage of the bootstrap intervals for the difference between fitted and observed joint rates.

\section{Numerical integration and simulation settings}\label{app:numerics}

This appendix describes the numerical calculations for the prior, posterior and joint survival probabilities, together with their accuracy checks. It also records the generating parameters and the denominators used in the simulation summaries.

\subsection{Working posterior and numerical integration}
We combine the rank-based pseudo-likelihood with a restricted Jeffreys prior based on the continuous copula model\citep{jeffreys1946} to form a working posterior. This construction treats the estimated marginal distributions as known and encounters as independent. Let
\begin{equation}\label{eq:prior}
\begin{split}
I(\boldsymbol\theta)&=\int_0^1\int_0^1
\boldsymbol s_{\boldsymbol\theta}(u,v)\boldsymbol s_{\boldsymbol\theta}(u,v)^{\mathsf T}
c_{\boldsymbol\theta}(u,v)\,du\,dv,\\
\pi_J(\boldsymbol\theta)&\ \propto\ |I(\boldsymbol\theta)|^{1/2}\mathbf 1(\boldsymbol\theta\in\Theta),
\end{split}
\end{equation}
where $\boldsymbol s_{\boldsymbol\theta}=\frac{\partial\log c_{\boldsymbol\theta}}{\partial\boldsymbol\theta}$ and $\Theta$ is the bounded domain. For the Student $t$ copula, $I$ is the full two-dimensional information matrix. The prior is defined under the continuous copula model, while empirical-margin estimation and patient clustering are accounted for through the uncertainty adjustment and resampling procedure.

The unadjusted posterior is
\begin{equation}\label{eq:workingposterior}
\pi_N(\boldsymbol\theta\mid\mathcal D)=
\frac{\exp\{\ell_N(\boldsymbol\theta)\}\pi_J(\boldsymbol\theta)}
{\int_\Theta\exp\{\ell_N(\boldsymbol t)\}\pi_J(\boldsymbol t)\,d\boldsymbol t}.
\end{equation}
Here $\mathcal D$ denotes the analysed sample. This distribution serves as the unadjusted comparator. For numerical integration, we work in transformed parameter coordinates and use trapezoidal grid weights. The prior is computed in these coordinates and so includes the Jacobian once. The grid approximation is compared with adaptive quadrature, requiring relative error in the normalising constant and absolute errors in scaled first and second moments and selected marginal cumulative probabilities to be at most $10^{-5}$.

\subsection{Coordinates and the prior}
We use $z=\operatorname{arctanh}(\rho)$ for Gaussian dependence, and $(\operatorname{arctanh}(\rho),\log\nu)$ for the Student $t$ parameters. The transformations are $z=\operatorname{arsinh}(\theta)$ for Frank, $z=\log(\theta-1)$ for Gumbel and Joe, and $z=\log\theta$ for survival Clayton. Each parameter domain is transformed using the map for its family.

To express the prior in these coordinates, let $J_{\theta z}=\partial\boldsymbol\theta/\partial\boldsymbol z^{\mathsf T}$. The Fisher information matrices satisfy
\begin{equation}\label{eq:jeffreyschange}
I_z=J_{\theta z}^{\mathsf T}I_\theta J_{\theta z},\qquad
|I_z|^{1/2}=|I_\theta|^{1/2}|\det J_{\theta z}|.
\end{equation}
We calculate information in the $\boldsymbol z$ coordinates, so $|I_z|^{1/2}$ already includes the Jacobian in equation~\eqref{eq:jeffreyschange}.

For the Gaussian copula, $z=\operatorname{arctanh}(\rho)$ gives $I_z=1+\rho^2$. Its restricted Jeffreys density in $z$ is therefore
\[
\pi_J(z)\propto\sqrt{1+\tanh^2 z}\,\mathbf 1\{z\in\mathcal Z\},
\]
where $\mathcal Z$ is the transformed parameter domain.

For the other families, we calculate expected score information by numerical integration under the continuous copula model. At the Frank independence limit, $|I_z(0)|^{1/2}=1/6$. Independent uniform coordinates are transformed into copula coordinates using an inverse conditional distribution. To concentrate integration nodes near zero and one, we use Gauss--Legendre integration after a logistic transformation. Cubic interpolation gives the log prior between tabulated values. We check these evaluations against reference calculations with a larger integration order and wider transformed range, using a relative tolerance of 0.5\%.

On a one-dimensional axis $z_1<\cdots<z_K$, the trapezoidal weights are
\begin{equation}\label{eq:weights}
w_1=(z_2-z_1)/2,\qquad
w_k=(z_{k+1}-z_{k-1})/2,\qquad
w_K=(z_K-z_{K-1})/2.
\end{equation}
The middle formula applies to $2\le k\le K-1$. For two parameters, the weights are products of the marginal weights. Each grid node receives mass proportional to
$\exp\{\ell_N(\boldsymbol z_k)\}|I_z(\boldsymbol z_k)|^{1/2}w_k$.
The grid is concentrated around the posterior mode. At each refinement, the dense local component of each axis uses 65, 129, 257, 513, 1,025, 2,049, 4,097 or 8,193 nodes, stopping at 1,025 for two parameters. The axis combines this component, extending up to eight local scales on either side of the mode within the domain, with a coarser grid spanning the full domain and the mode itself. Each local scale is the square root of a diagonal entry of the inverse posterior curvature, capped at one quarter of the domain width; if the curvature is not positive definite, we use one quarter of the domain width. Duplicate nodes are removed. The grid normalising constant, first and second raw moments of each coordinate rescaled to $[0,1]$, and a marginal cumulative probability at a grid node near each posterior mean are compared with adaptive quadrature. The relative error in the normalising constant and the absolute errors in the scaled moments and cumulative probabilities must each be at most $10^{-5}$. This check does not verify every posterior quantile or interval endpoint. We draw from the continuous piecewise-linear density in one dimension or the bilinear density in two dimensions. Application posterior summaries use 4,000 draws, while the eight-component randomised-rank mixture uses 500 draws from each component.

For the six main families on the main parameter ranges, we obtain the maximum pseudo-likelihood estimate by bounded L-BFGS-B optimisation, switching to bounded Powell optimisation if L-BFGS-B does not report convergence. The posterior mode is located by bounded L-BFGS-B optimisation, starting from the maximum pseudo-likelihood estimate. The minimised objective for this search is the negative log unnormalised posterior in $\boldsymbol z$ coordinates, namely minus the sum of the log pseudo-likelihood and the log prior in those coordinates. In the simulation and sensitivity analyses, we retry the posterior-mode search with bounded Powell optimisation if L-BFGS-B does not report convergence or returns a non-finite objective. Powell starts from the final L-BFGS-B point when all its coordinates are finite and from the original point otherwise. The retry is accepted only when it reports convergence and returns a finite objective no greater than the value at the original L-BFGS-B starting point plus $10^{-8}$. The posterior must also satisfy the quadrature accuracy checks.

For BB1, we minimise the negative average log pseudo-likelihood, weighting each distinct pseudo-observation pair by its count. The search starts at $(0.05,1.05)$ and also at the pyvinecopulib weighted maximum-likelihood estimate when that estimate is finite. If the library fit raises an exception, we use $(0,1.05)$ as the additional start. Each start uses bounded L-BFGS-B, with bounded Powell from that same original start if L-BFGS-B fails to converge or returns a non-finite objective. The best converged finite result is retained, and its objective must be no greater than the best starting objective plus $10^{-10}$. BB7 instead uses pyvinecopulib's built-in weighted maximum-likelihood fit. These two likelihood-based sensitivity analyses do not use a posterior objective.

\Needspace{10\baselineskip}
\subsection{Joint survival probabilities}
To calculate the probability that both variables exceed fixed percentile cut-offs $a,b$, we use the copula survival function
\begin{equation}\label{eq:survival}
S_{\boldsymbol\theta}(a,b)=1-a-b+C_{\boldsymbol\theta}(a,b)
=\int_a^1\left\{1-\frac{\partial C_{\boldsymbol\theta}(u,b)}{\partial u}\right\}\,du.
\end{equation}
For Gaussian and Student $t$ copulas, we evaluate the integral on the original normal or Student scale. With $x_a=\Phi^{-1}(a)$ and $y_b=\Phi^{-1}(b)$, the Gaussian expression is
\begin{equation}\label{eq:normalintegral}
S_\rho(a,b)=\int_{x_a}^{\infty}\phi(x)
\Phi\left(\frac{\rho x-y_b}{\sqrt{1-\rho^2}}\right)\,dx.
\end{equation}
For Student $t$, set $x_a=t_\nu^{-1}(a)$ and $y_b=t_\nu^{-1}(b)$. If $f_\nu$ denotes the univariate density, then
\begin{equation}\label{eq:tintegral}
S_{\rho,\nu}(a,b)=\int_{x_a}^{\infty}f_\nu(x)
t_{\nu+1}\left(\frac{\rho x-y_b}
{\sqrt{(\nu+x^2)(1-\rho^2)/(\nu+1)}}\right)\,dx.
\end{equation}
For the remaining main families, we use the conditional-distribution form in equation~\eqref{eq:survival}. For BB1 and BB7, we use their closed-form distribution functions. The absolute and relative quadrature tolerances are $10^{-11}$ and $10^{-9}$, respectively. Independent reference checks use a tolerance of $10^{-7}$. In the matched-family simulations, the five families other than Student $t$ used the copula distribution function directly. Student $t$ used the integral in equation~\eqref{eq:tintegral}.

\subsection{Matched-family simulation targets}
The generating parameters are $\rho=0.45$ for Gaussian and $(\rho,\nu)=(0.45,4)$ for Student $t$. For the other families, we use $\theta=3$ for Frank, $\theta=1.7$ for Gumbel, $\theta=2$ for Joe and $\theta=1.2$ for survival Clayton. The fitted family matches the generating family. The targets are the probabilities in equation~\eqref{eq:nominal} at $q=0.90$ and $q=0.95$ for the generating continuous copula. We retain these targets in the rounded setting. Each unrounded observation has the stated generating copula, but rounding can change the population target of the mid-rank fit.

For coverage among available intervals, the denominator is the number of available intervals. For coverage counting failures as misses, the denominator is all 2,000 replicates. Each Monte Carlo standard error uses the same denominator as its corresponding coverage estimate. Error summaries use available estimates.

\section{Detailed simulation results}\label{app:simulation-details}
\begin{table}[!htbp]
\caption{Coverage (\%) of nominal 95\% intervals.}\label{tab:simulation-coverage}
\centering\small
\setlength{\tabcolsep}{4pt}
\begin{tabular}{@{}llrrrr@{}}
\toprule
Family & Percentile & \shortstack{Uncorrected\\posterior} & \shortstack{Sandwich-adjusted\\posterior} & \shortstack{Patient-cluster\\bootstrap} & \shortstack{Random-tie\\posterior}\\
\midrule
\multicolumn{6}{@{}l}{\textit{Independent continuous observations}}\\[2pt]
Gaussian & 90 and 95 & 92.20 (0.60) & 94.80 (0.50) & 94.45 (0.51) & 92.35 (0.59)\\
Student $t$ & 90 & 92.45 (0.59) & 94.75 (0.50) & 94.90 (0.49) & 92.55 (0.59)\\
 & 95 & 91.50 (0.62) & 94.00 (0.53) & 94.40 (0.51) & 91.50 (0.62)\\
Frank & 90 and 95 & 93.60 (0.55) & 94.80 (0.50) & 94.50 (0.51) & 93.60 (0.55)\\
Gumbel & 90 and 95 & 89.75 (0.68) & 95.35 (0.47) & 95.70 (0.45) & 89.75 (0.68)\\
Joe & 90 and 95 & 85.60 (0.79) & 94.05 (0.53) & 94.10 (0.53) & 85.85 (0.78)\\
Survival Clayton & 90 and 95 & 85.35 (0.79) & 94.20 (0.52) & 94.60 (0.51) & 85.45 (0.79)\\
\addlinespace
\multicolumn{6}{@{}l}{\textit{Rounded observations with repeated measurements}}\\[2pt]
Gaussian & 90 and 95 & 86.90 (0.75) & 94.70 (0.50) & 94.70 (0.50) & 87.40 (0.74)\\
Student $t$ & 90 & 86.15 (0.77) & 94.55 (0.51) & 94.30 (0.52) & 86.85 (0.76)\\
 & 95 & 85.95 (0.78) & 94.20 (0.52) & 94.60 (0.51) & 87.00 (0.75)\\
Frank & 90 and 95 & 88.60 (0.71) & 94.70 (0.50) & 94.50 (0.51) & 88.60 (0.71)\\
Gumbel & 90 and 95 & 82.05 (0.86) & 94.35 (0.52) & 94.55 (0.51) & 82.40 (0.85)\\
Joe & 90 and 95 & 81.45 (0.87) & 94.85 (0.49) & 94.90 (0.49) & 81.85 (0.86)\\
Survival Clayton & 90 and 95 & 79.90 (0.90) & 94.45 (0.51) & 94.90 (0.49) & 79.70 (0.90)\\
\bottomrule
\end{tabular}
\end{table}

For the five one-parameter families, $R_q$ increases with the copula parameter at both percentiles, so both transformations preserve the order of the parameter draws. Consequently, for each of these five families and each method, the intervals covered their targets in the same replicates at both percentiles. Rows labelled 90 and 95 in the table therefore have identical coverage at the two percentiles.

The sandwich-adjusted posterior and the patient-cluster bootstrap gave wider intervals on average than the uncorrected posterior for every family, setting and percentile. For the Student $t$ copula at the 95th percentile with rounded, repeated observations, coverage was 85.95\% for the uncorrected posterior, 94.20\% for the sandwich-adjusted posterior and 94.60\% for the patient-cluster bootstrap. The corresponding mean interval widths were 4.89, 6.11 and 6.48 per 1,000, respectively.

Only the Student $t$ fits had unavailable intervals. At each percentile, the sandwich-adjusted posterior interval was unavailable in two replicates with continuous observations and seven with rounded, repeated observations. In all nine of these replicates, the estimate lay at a parameter bound. In one of the rounded replicates, the posterior grids also failed the accuracy check, so the uncorrected and random-tie posterior intervals were unavailable there as well. All patient-cluster bootstrap intervals were available.

Tables~\ref{tab:sim-detail-c90}--\ref{tab:sim-detail-r95} give results for each family and interval method. Bias, root mean squared error and mean interval width use the replicates with available intervals and are on the probability scale multiplied by 1,000. Coverage is the percentage of available intervals containing the true probability. In contrast, Section~\ref{sec:simulation-results} and Table~\ref{tab:simulation-coverage} report coverage over all 2,000 replicates per family and setting.

The unavailable column counts replicates without an interval for the method shown. The boundary column counts original-sample fits at a parameter bound. Both columns give counts out of 2,000 replicates, and the boundary count is repeated across methods for the same family and setting. The sandwich-adjusted posterior uses the empirical-margin and patient-cluster adjustment in Section~\ref{sec:calibration}. The patient-cluster bootstrap resamples whole patients.

\begin{table}[!htbp]
\caption{Independent continuous pairs at the 90th percentile.}\label{tab:sim-detail-c90}
\centering\small
\setlength{\tabcolsep}{3.0pt}
\begin{tabular}{@{}llrrrrrr@{}}
\toprule
Family & Method & Bias & \shortstack{Root mean\\squared error} & \shortstack{Mean\\width} & \shortstack{Coverage\\(\%)} & Unavailable & Boundary\\
\midrule
Gaussian & Uncorrected posterior & 0.227 & 2.086 & 7.272 & 92.20 & 0 & 0\\
 & Sandwich-adjusted posterior & 0.274 & 2.091 & 7.884 & 94.80 & 0 & 0\\
 & Patient-cluster bootstrap & 0.257 & 2.090 & 7.893 & 94.45 & 0 & 0\\
 & Random-tie posterior & 0.226 & 2.087 & 7.264 & 92.35 & 0 & 0\\
\addlinespace[2pt]
Student $t$ & Uncorrected posterior & 0.206 & 2.465 & 8.652 & 92.45 & 0 & 2\\
 & Sandwich-adjusted posterior & 0.250 & 2.422 & 9.350 & 94.84 & 2 & 2\\
 & Patient-cluster bootstrap & 0.294 & 2.483 & 9.624 & 94.90 & 0 & 2\\
 & Random-tie posterior & 0.209 & 2.465 & 8.649 & 92.55 & 0 & 2\\
\addlinespace[2pt]
Frank & Uncorrected posterior & 0.019 & 1.511 & 5.608 & 93.60 & 0 & 0\\
 & Sandwich-adjusted posterior & 0.066 & 1.509 & 5.816 & 94.80 & 0 & 0\\
 & Patient-cluster bootstrap & 0.035 & 1.512 & 5.840 & 94.50 & 0 & 0\\
 & Random-tie posterior & 0.020 & 1.512 & 5.607 & 93.60 & 0 & 0\\
\addlinespace[2pt]
Gumbel & Uncorrected posterior & 0.241 & 2.323 & 7.610 & 89.75 & 0 & 0\\
 & Sandwich-adjusted posterior & 0.305 & 2.323 & 9.084 & 95.35 & 0 & 0\\
 & Patient-cluster bootstrap & 0.265 & 2.324 & 9.116 & 95.70 & 0 & 0\\
 & Random-tie posterior & 0.242 & 2.324 & 7.613 & 89.75 & 0 & 0\\
\addlinespace[2pt]
Joe & Uncorrected posterior & 0.144 & 2.797 & 8.238 & 85.60 & 0 & 0\\
 & Sandwich-adjusted posterior & 0.214 & 2.788 & 10.637 & 94.05 & 0 & 0\\
 & Patient-cluster bootstrap & 0.175 & 2.796 & 10.715 & 94.10 & 0 & 0\\
 & Random-tie posterior & 0.145 & 2.795 & 8.239 & 85.85 & 0 & 0\\
\addlinespace[2pt]
Survival Clayton & Uncorrected posterior & 0.210 & 2.837 & 8.361 & 85.35 & 0 & 0\\
 & Sandwich-adjusted posterior & 0.284 & 2.832 & 10.919 & 94.20 & 0 & 0\\
 & Patient-cluster bootstrap & 0.243 & 2.837 & 11.024 & 94.60 & 0 & 0\\
 & Random-tie posterior & 0.211 & 2.837 & 8.353 & 85.45 & 0 & 0\\
\bottomrule
\end{tabular}
\end{table}

\begin{table}[!htbp]
\caption{Independent continuous pairs at the 95th percentile.}\label{tab:sim-detail-c95}
\centering\small
\setlength{\tabcolsep}{3.0pt}
\begin{tabular}{@{}llrrrrrr@{}}
\toprule
Family & Method & Bias & \shortstack{Root mean\\squared error} & \shortstack{Mean\\width} & \shortstack{Coverage\\(\%)} & Unavailable & Boundary\\
\midrule
Gaussian & Uncorrected posterior & 0.118 & 1.004 & 3.492 & 92.20 & 0 & 0\\
 & Sandwich-adjusted posterior & 0.141 & 1.007 & 3.788 & 94.80 & 0 & 0\\
 & Patient-cluster bootstrap & 0.133 & 1.006 & 3.794 & 94.45 & 0 & 0\\
 & Random-tie posterior & 0.118 & 1.004 & 3.488 & 92.35 & 0 & 0\\
\addlinespace[2pt]
Student $t$ & Uncorrected posterior & 0.155 & 1.414 & 4.891 & 91.50 & 0 & 2\\
 & Sandwich-adjusted posterior & 0.171 & 1.369 & 5.225 & 94.09 & 2 & 2\\
 & Patient-cluster bootstrap & 0.207 & 1.435 & 5.514 & 94.40 & 0 & 2\\
 & Random-tie posterior & 0.156 & 1.414 & 4.889 & 91.50 & 0 & 2\\
\addlinespace[2pt]
Frank & Uncorrected posterior & 0.010 & 0.493 & 1.829 & 93.60 & 0 & 0\\
 & Sandwich-adjusted posterior & 0.025 & 0.493 & 1.898 & 94.80 & 0 & 0\\
 & Patient-cluster bootstrap & 0.015 & 0.494 & 1.906 & 94.50 & 0 & 0\\
 & Random-tie posterior & 0.010 & 0.494 & 1.829 & 93.60 & 0 & 0\\
\addlinespace[2pt]
Gumbel & Uncorrected posterior & 0.126 & 1.227 & 4.019 & 89.75 & 0 & 0\\
 & Sandwich-adjusted posterior & 0.160 & 1.226 & 4.797 & 95.35 & 0 & 0\\
 & Patient-cluster bootstrap & 0.139 & 1.227 & 4.814 & 95.70 & 0 & 0\\
 & Random-tie posterior & 0.127 & 1.227 & 4.021 & 89.75 & 0 & 0\\
\addlinespace[2pt]
Joe & Uncorrected posterior & 0.070 & 1.442 & 4.252 & 85.60 & 0 & 0\\
 & Sandwich-adjusted posterior & 0.106 & 1.437 & 5.491 & 94.05 & 0 & 0\\
 & Patient-cluster bootstrap & 0.086 & 1.442 & 5.532 & 94.10 & 0 & 0\\
 & Random-tie posterior & 0.070 & 1.441 & 4.253 & 85.85 & 0 & 0\\
\addlinespace[2pt]
Survival Clayton & Uncorrected posterior & 0.105 & 1.539 & 4.539 & 85.35 & 0 & 0\\
 & Sandwich-adjusted posterior & 0.145 & 1.535 & 5.927 & 94.20 & 0 & 0\\
 & Patient-cluster bootstrap & 0.123 & 1.538 & 5.986 & 94.60 & 0 & 0\\
 & Random-tie posterior & 0.106 & 1.538 & 4.535 & 85.45 & 0 & 0\\
\bottomrule
\end{tabular}
\end{table}

\begin{table}[!htbp]
\caption{Rounded, repeated pairs at the 90th percentile.}\label{tab:sim-detail-r90}
\centering\small
\setlength{\tabcolsep}{3.0pt}
\begin{tabular}{@{}llrrrrrr@{}}
\toprule
Family & Method & Bias & \shortstack{Root mean\\squared error} & \shortstack{Mean\\width} & \shortstack{Coverage\\(\%)} & Unavailable & Boundary\\
\midrule
Gaussian & Uncorrected posterior & 0.396 & 2.426 & 7.278 & 86.90 & 0 & 0\\
 & Sandwich-adjusted posterior & 0.446 & 2.435 & 9.271 & 94.70 & 0 & 0\\
 & Patient-cluster bootstrap & 0.427 & 2.432 & 9.294 & 94.70 & 0 & 0\\
 & Random-tie posterior & 0.308 & 2.405 & 7.270 & 87.40 & 0 & 0\\
\addlinespace[2pt]
Student $t$ & Uncorrected posterior & 0.337 & 2.934 & 8.643 & 86.19 & 1 & 7\\
 & Sandwich-adjusted posterior & 0.347 & 2.854 & 11.019 & 94.88 & 7 & 7\\
 & Patient-cluster bootstrap & 0.424 & 2.963 & 11.362 & 94.30 & 0 & 7\\
 & Random-tie posterior & 0.193 & 2.915 & 8.657 & 86.89 & 1 & 7\\
\addlinespace[2pt]
Frank & Uncorrected posterior & 0.054 & 1.773 & 5.605 & 88.60 & 0 & 0\\
 & Sandwich-adjusted posterior & 0.106 & 1.769 & 6.870 & 94.70 & 0 & 0\\
 & Patient-cluster bootstrap & 0.068 & 1.774 & 6.911 & 94.50 & 0 & 0\\
 & Random-tie posterior & 0.014 & 1.769 & 5.608 & 88.60 & 0 & 0\\
\addlinespace[2pt]
Gumbel & Uncorrected posterior & 0.125 & 2.786 & 7.642 & 82.05 & 0 & 0\\
 & Sandwich-adjusted posterior & 0.199 & 2.779 & 10.712 & 94.35 & 0 & 0\\
 & Patient-cluster bootstrap & 0.151 & 2.784 & 10.778 & 94.55 & 0 & 0\\
 & Random-tie posterior & -0.026 & 2.782 & 7.678 & 82.40 & 0 & 0\\
\addlinespace[2pt]
Joe & Uncorrected posterior & 0.137 & 3.148 & 8.248 & 81.45 & 0 & 0\\
 & Sandwich-adjusted posterior & 0.217 & 3.141 & 12.493 & 94.85 & 0 & 0\\
 & Patient-cluster bootstrap & 0.170 & 3.149 & 12.598 & 94.90 & 0 & 0\\
 & Random-tie posterior & -0.053 & 3.143 & 8.298 & 81.85 & 0 & 0\\
\addlinespace[2pt]
Survival Clayton & Uncorrected posterior & -0.069 & 3.324 & 8.421 & 79.90 & 0 & 0\\
 & Sandwich-adjusted posterior & 0.012 & 3.306 & 12.833 & 94.45 & 0 & 0\\
 & Patient-cluster bootstrap & -0.035 & 3.320 & 12.938 & 94.90 & 0 & 0\\
 & Random-tie posterior & -0.277 & 3.332 & 8.487 & 79.70 & 0 & 0\\
\bottomrule
\end{tabular}
\end{table}

\begin{table}[!htbp]
\caption{Rounded, repeated pairs at the 95th percentile.}\label{tab:sim-detail-r95}
\centering\small
\setlength{\tabcolsep}{3.0pt}
\begin{tabular}{@{}llrrrrrr@{}}
\toprule
Family & Method & Bias & \shortstack{Root mean\\squared error} & \shortstack{Mean\\width} & \shortstack{Coverage\\(\%)} & Unavailable & Boundary\\
\midrule
Gaussian & Uncorrected posterior & 0.203 & 1.170 & 3.499 & 86.90 & 0 & 0\\
 & Sandwich-adjusted posterior & 0.227 & 1.175 & 4.457 & 94.70 & 0 & 0\\
 & Patient-cluster bootstrap & 0.218 & 1.173 & 4.473 & 94.70 & 0 & 0\\
 & Random-tie posterior & 0.160 & 1.158 & 3.493 & 87.40 & 0 & 0\\
\addlinespace[2pt]
Student $t$ & Uncorrected posterior & 0.210 & 1.667 & 4.888 & 85.99 & 1 & 7\\
 & Sandwich-adjusted posterior & 0.203 & 1.588 & 6.112 & 94.53 & 7 & 7\\
 & Patient-cluster bootstrap & 0.261 & 1.698 & 6.476 & 94.60 & 0 & 7\\
 & Random-tie posterior & 0.118 & 1.654 & 4.898 & 87.04 & 1 & 7\\
\addlinespace[2pt]
Frank & Uncorrected posterior & 0.023 & 0.579 & 1.829 & 88.60 & 0 & 0\\
 & Sandwich-adjusted posterior & 0.040 & 0.578 & 2.243 & 94.70 & 0 & 0\\
 & Patient-cluster bootstrap & 0.027 & 0.579 & 2.257 & 94.50 & 0 & 0\\
 & Random-tie posterior & 0.010 & 0.577 & 1.829 & 88.60 & 0 & 0\\
\addlinespace[2pt]
Gumbel & Uncorrected posterior & 0.065 & 1.471 & 4.036 & 82.05 & 0 & 0\\
 & Sandwich-adjusted posterior & 0.104 & 1.468 & 5.657 & 94.35 & 0 & 0\\
 & Patient-cluster bootstrap & 0.078 & 1.470 & 5.692 & 94.55 & 0 & 0\\
 & Random-tie posterior & -0.015 & 1.469 & 4.056 & 82.40 & 0 & 0\\
\addlinespace[2pt]
Joe & Uncorrected posterior & 0.065 & 1.623 & 4.258 & 81.45 & 0 & 0\\
 & Sandwich-adjusted posterior & 0.106 & 1.619 & 6.452 & 94.85 & 0 & 0\\
 & Patient-cluster bootstrap & 0.082 & 1.623 & 6.507 & 94.90 & 0 & 0\\
 & Random-tie posterior & -0.033 & 1.622 & 4.286 & 81.85 & 0 & 0\\
\addlinespace[2pt]
Survival Clayton & Uncorrected posterior & -0.049 & 1.805 & 4.577 & 79.90 & 0 & 0\\
 & Sandwich-adjusted posterior & -0.006 & 1.795 & 6.975 & 94.45 & 0 & 0\\
 & Patient-cluster bootstrap & -0.031 & 1.803 & 7.033 & 94.90 & 0 & 0\\
 & Random-tie posterior & -0.162 & 1.812 & 4.616 & 79.70 & 0 & 0\\
\bottomrule
\end{tabular}
\end{table}

\FloatBarrier
\section{Additional results for the laboratory data}\label{app:application-details}

\begin{figure}[!htbp]
\centering
   \includegraphics[width=0.95\linewidth]{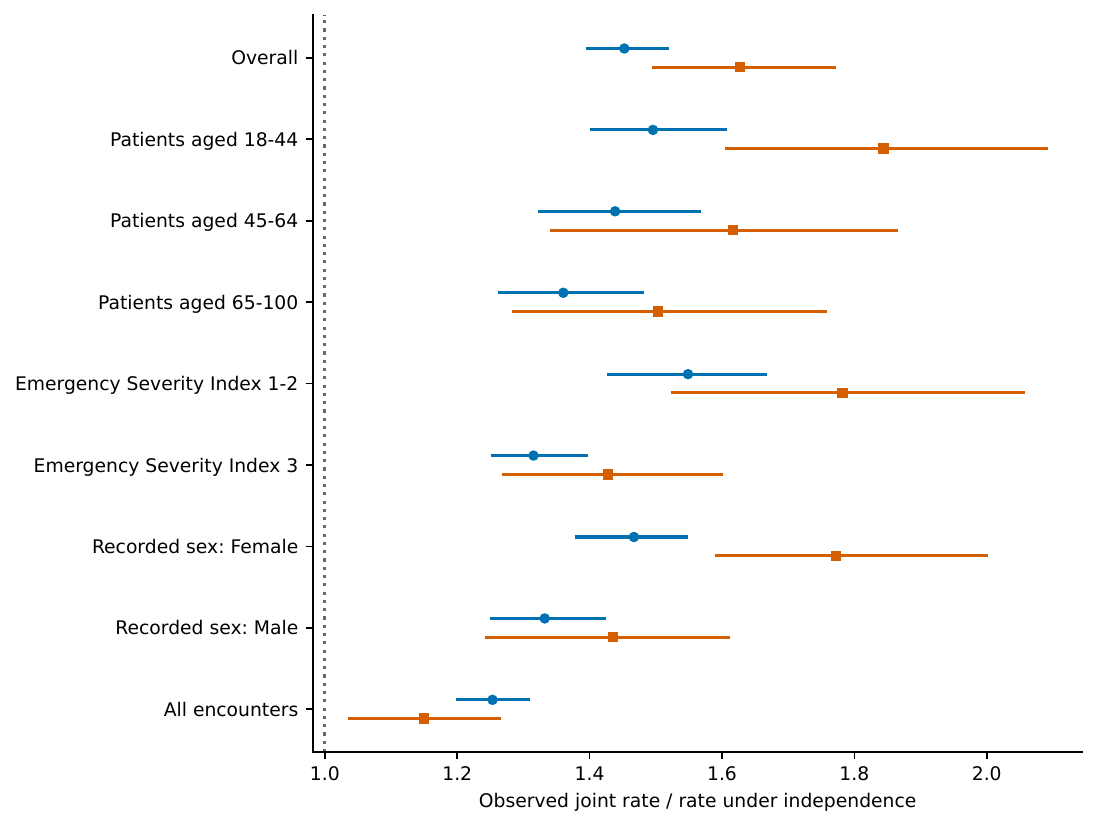}
   \caption{Joint exceedance ratios relative to independence at the 90th (circles) and 95th (squares) percentiles.}\label{fig:app-ratios}
\par\noindent
\textbf{Alt text:} Both intervals lie above one in every row. At the 95th percentile the ratio is 1.63 for first encounters, 1.84 for patients aged 18--44 years and 1.15 for all encounters.
\end{figure}

Table~\ref{tab:app-groups} gives the group sizes and empirical results at both percentiles for the first qualifying encounter. Each group has its own cut-offs, and confidence intervals use 999 patient resamples. Entries marked \textsuperscript{$\dagger$} are descriptive estimates for groups with fewer than 50 joint exceedances at the 95th percentile. Counts below 11, and estimates derived from them, are suppressed.

The sensitivity to prior bounds is shown in Table~\ref{tab:app-prior}, which compares sandwich-adjusted posterior estimates at nominal percentiles under the two prior supports. Values are posterior medians per 1,000 with central 95\% intervals for the first-encounter sample. Table~\ref{tab:support} gives the parameter bounds.

To examine the effects of ties and repeated encounters, Table~\ref{tab:app-ties} compares fitted probabilities at the observed cut-offs under the alternative rules. Each sample uses its own observed cut-offs. Parentheses give pointwise 95\% confidence intervals from 999 patient resamples. Random tie-breaking averages fitted probabilities over eight orderings.

The paired comparison in Table~\ref{tab:app-visits} uses the same 999 patient resamples for all columns, with differences calculated as all encounters minus first encounters. Parentheses give pointwise 95\% confidence intervals. Table~\ref{tab:app-empirical} uses 2,000 resamples for the separate sample estimates.

\begin{table}[!htbp]
\caption{Joint exceedances by patient group at the first qualifying encounter.}\label{tab:app-groups}
\centering\small
\setlength{\tabcolsep}{4pt}
\begin{tabular}{@{}lrrrrrr@{}}
\toprule
Group & Patients & Percentile & \shortstack{Creatinine\\(mg/dL)} & \shortstack{AST\\(U/L)} & Joint count & \shortstack{Ratio to independence\\(95\% CI)}\\
\midrule
\multicolumn{7}{@{}l}{\textit{Age (years)}}\\
18--44 & 41,323 & 90 & 1.06 & 55 & 604 & 1.50 (1.40, 1.61)\\
 &  & 95 & 1.19 & 83 & 183 & 1.84 (1.60, 2.09)\\
45--64 & 30,946 & 90 & 1.27 & 65 & 432 & 1.44 (1.32, 1.57)\\
 &  & 95 & 1.68 & 102 & 122 & 1.62 (1.34, 1.87)\\
65--100 & 35,118 & 90 & 1.64 & 59 & 473 & 1.36 (1.26, 1.48)\\
 &  & 95 & 2.25 & 88 & 130 & 1.50 (1.28, 1.76)\\
\addlinespace[2pt]
\multicolumn{7}{@{}l}{\textit{Emergency Severity Index}}\\
1--2 & 31,237 & 90 & 1.45 & 72 & 470 & 1.55 (1.43, 1.67)\\
 &  & 95 & 1.97 & 116 & 136 & 1.78 (1.52, 2.06)\\
3 & 72,781 & 90 & 1.24 & 55 & 925 & 1.32 (1.25, 1.40)\\
 &  & 95 & 1.60 & 81 & 250 & 1.43 (1.27, 1.60)\\
4--5 & 2,911 & 90 & 1.10 & 46 & 30 & 1.07\textsuperscript{$\dagger$}\\
 &  & 95 & 1.24 & 61 & $<$11 & --\\
Missing & 458 & 90 & 1.44 & 80 & 11 & 2.49\textsuperscript{$\dagger$}\\
 &  & 95 & 1.91 & 180 & $<$11 & --\\
\addlinespace[2pt]
\multicolumn{7}{@{}l}{\textit{Recorded sex}}\\
Female & 58,445 & 90 & 1.06 & 53 & 842 & 1.47 (1.38, 1.55)\\
 &  & 95 & 1.38 & 79 & 254 & 1.77 (1.59, 2.00)\\
Male & 48,919 & 90 & 1.48 & 66 & 642 & 1.33 (1.25, 1.42)\\
 &  & 95 & 2.01 & 104 & 175 & 1.44 (1.24, 1.61)\\
Unknown & 23 & 90 & -- & -- & $<$11 & --\\
 &  & 95 & -- & -- & $<$11 & --\\
\bottomrule
\end{tabular}
\end{table}

\begin{table}[!htbp]
\caption{Sandwich-adjusted posterior joint probabilities per 1,000 under two prior supports.}\label{tab:app-prior}
\centering\small
\setlength{\tabcolsep}{4pt}
\begin{tabular}{@{}lrrr@{}}
\toprule
Family & Percentile & Main support & Wider support\\
\midrule
Gaussian & 90 & 12.313 (12.085, 12.528) & 12.312 (12.082, 12.535)\\
 & 95 & 3.329 (3.244, 3.409) & 3.328 (3.244, 3.411)\\
Student $t$ & 90 & 15.793 (15.475, 16.094) & 15.789 (15.493, 16.110)\\
 & 95 & 5.405 (5.240, 5.563) & 5.403 (5.246, 5.569)\\
Frank & 90 & 12.442 (12.268, 12.612) & 12.440 (12.267, 12.622)\\
 & 95 & 3.191 (3.142, 3.240) & 3.191 (3.141, 3.243)\\
Gumbel & 90 & 15.420 (14.991, 15.863) & 15.427 (14.995, 15.865)\\
 & 95 & 5.435 (5.203, 5.674) & 5.439 (5.205, 5.675)\\
Joe & 90 & 14.970 (14.468, 15.485) & 14.970 (14.471, 15.481)\\
 & 95 & 5.378 (5.089, 5.674) & 5.378 (5.091, 5.672)\\
Survival Clayton & 90 & 13.819 (13.425, 14.247) & 13.821 (13.410, 14.247)\\
 & 95 & 4.229 (4.042, 4.435) & 4.231 (4.035, 4.435)\\
\bottomrule
\end{tabular}
\end{table}

\begin{table}[!htbp]
\caption{Fitted joint rates per 1,000 at the observed laboratory cut-offs.}\label{tab:app-ties}
\centering\small
\setlength{\tabcolsep}{4pt}
\begin{tabular}{@{}lrrrr@{}}
\toprule
Family & Percentile & \shortstack{First encounter\\Mid-ranks} & \shortstack{First encounter\\Random tie-breaking} & \shortstack{All encounters\\Mid-ranks}\\
\midrule
Gaussian & 90 & 12.19 (11.71, 12.36) & 12.19 (11.71, 12.33) & 10.84 (10.51, 11.19)\\
 & 95 & 3.28 (3.20, 3.37) & 3.28 (3.20, 3.37) & 2.86 (2.74, 2.95)\\
Student $t$ & 90 & 15.66 (15.07, 15.89) & 15.64 (15.05, 15.85) & 14.32 (13.94, 14.74)\\
 & 95 & 5.34 (5.18, 5.50) & 5.33 (5.17, 5.50) & 4.89 (4.69, 5.05)\\
Frank & 90 & 12.32 (11.84, 12.44) & 12.31 (11.83, 12.44) & 11.27 (11.00, 11.57)\\
 & 95 & 3.14 (3.09, 3.21) & 3.14 (3.09, 3.21) & 2.91 (2.83, 2.96)\\
Gumbel & 90 & 15.29 (14.62, 15.61) & 15.29 (14.63, 15.60) & 12.60 (12.02, 13.16)\\
 & 95 & 5.37 (5.15, 5.62) & 5.37 (5.16, 5.61) & 4.02 (3.69, 4.30)\\
Joe & 90 & 14.84 (14.14, 15.20) & 14.84 (14.11, 15.22) & 11.97 (11.39, 12.56)\\
 & 95 & 5.31 (5.04, 5.61) & 5.31 (5.05, 5.61) & 3.76 (3.42, 4.07)\\
Survival Clayton & 90 & 13.70 (13.06, 14.01) & 13.70 (13.05, 13.99) & 11.43 (10.93, 11.92)\\
 & 95 & 4.17 (3.98, 4.39) & 4.17 (3.99, 4.37) & 3.21 (2.97, 3.40)\\
\bottomrule
\end{tabular}
\end{table}

\begin{table}[!htbp]
\caption{Ratios to independence under alternative encounter and cut-off rules.}\label{tab:app-visits}
\centering\small
\setlength{\tabcolsep}{4pt}
\begin{tabular}{@{}rrrrr@{}}
\toprule
Percentile & \shortstack{First encounters\\Own cut-offs} & \shortstack{All encounters\\Own cut-offs} & \shortstack{All encounters\\First-encounter cut-offs} & \shortstack{Paired difference\\Own cut-offs}\\
\midrule
90 & 1.45 (1.39, 1.52) & 1.25 (1.20, 1.31) & 1.24 (1.19, 1.28) & $-$0.20 ($-$0.27, $-$0.14)\\
95 & 1.63 (1.49, 1.78) & 1.15 (1.04, 1.26) & 1.26 (1.17, 1.34) & $-$0.48 ($-$0.62, $-$0.34)\\
\bottomrule
\end{tabular}
\end{table}

\FloatBarrier
Table~\ref{tab:app-model-checks} reports all 64 family and group fits, including the pseudo-BIC, parameter-bound status and fitted-minus-observed gaps at both observed cut-offs. Gaps are per 1,000 patients, with pointwise 95\% intervals from 999 paired patient resamples. Positive gaps indicate overestimation. Pseudo-BIC comparisons are descriptive within a group and common rank representation. The ``At bound'' flag identifies fits within $10^{-5}$ of a parameter bound after division by the fitted domain width, using transformed coordinates for the six main families and original parameter coordinates for BB1 and BB7. Nominal interval coverage is not established for these fits. BB1 and BB7 are likelihood-based sensitivity comparisons. ESI denotes the Emergency Severity Index; Female and Male refer to recorded sex.

\begingroup
\small
\setlength{\tabcolsep}{2pt}
\begin{longtable}{@{}>{\raggedright\arraybackslash}p{2.65cm}>{\raggedright\arraybackslash}p{2.25cm}r>{\centering\arraybackslash}p{1.1cm}>{\raggedright\arraybackslash}p{3.0cm}>{\raggedright\arraybackslash}p{3.0cm}@{}}
\caption{Copula fit checks across families and patient groups.}\label{tab:app-model-checks}\\
\toprule
Group & Family & Pseudo-BIC & At bound & 90th gap per 1,000 (95\% CI) & 95th gap per 1,000 (95\% CI)\\
\midrule
\endfirsthead
\multicolumn{6}{@{}l}{\tablename~\thetable\ (continued)}\\
\toprule
Group & Family & Pseudo-BIC & At bound & 90th gap per 1,000 (95\% CI) & 95th gap per 1,000 (95\% CI)\\
\midrule
\endhead
Overall & Gaussian & -528.0 & No & -2.183 (-2.737, -1.665) & -0.726 (-1.084, -0.409)\\
Overall & Student $t$ & -1,687.1 & No & 1.284 (0.712, 1.803) & 1.335 (0.976, 1.642)\\
Overall & Frank & -918.6 & No & -2.060 (-2.651, -1.533) & -0.863 (-1.245, -0.550)\\
Overall & Gumbel & -701.6 & No & 0.915 (0.398, 1.394) & 1.368 (1.037, 1.671)\\
Overall & Joe & -449.8 & No & 0.466 (-0.002, 0.895) & 1.310 (1.012, 1.616)\\
Overall & Survival Clayton & -426.1 & No & -0.681 (-1.191, -0.208) & 0.168 (-0.140, 0.477)\\
Age 18--44 & Gaussian & -767.7 & No & -0.220 (-1.165, 0.664) & -0.339 (-0.924, 0.216)\\
Age 18--44 & Student $t$ & -1,094.7 & No & 2.810 (1.845, 3.616) & 1.487 (0.938, 2.015)\\
Age 18--44 & Frank & -1,094.1 & No & -0.503 (-1.481, 0.367) & -0.804 (-1.398, -0.241)\\
Age 18--44 & Gumbel & -759.7 & No & 4.489 (3.558, 5.263) & 2.976 (2.453, 3.481)\\
Age 18--44 & Joe & -477.2 & No & 3.880 (3.068, 4.596) & 2.957 (2.469, 3.477)\\
Age 18--44 & Survival Clayton & -570.5 & No & 2.742 (1.843, 3.503) & 1.537 (1.032, 2.061)\\
Age 45--64 & Gaussian & -9.2 & No & -3.481 (-4.650, -2.436) & -1.233 (-1.828, -0.589)\\
Age 45--64 & Student $t$ & -312.5 & No & -0.224 (-1.362, 0.750) & 0.652 (0.056, 1.276)\\
Age 45--64 & Frank & -59.3 & No & -3.078 (-4.262, -2.003) & -1.173 (-1.788, -0.520)\\
Age 45--64 & Gumbel & -54.0 & No & -1.348 (-2.344, -0.480) & 0.073 (-0.468, 0.629)\\
Age 45--64 & Joe & -37.0 & No & -1.414 (-2.305, -0.655) & 0.145 (-0.371, 0.660)\\
Age 45--64 & Survival Clayton & -17.0 & No & -2.596 (-3.576, -1.707) & -0.778 (-1.335, -0.207)\\
Age 65--100 & Gaussian & 1.0 & No & -4.064 (-5.083, -3.123) & -1.409 (-2.009, -0.864)\\
Age 65--100 & Student $t$ & -277.4 & No & -1.429 (-2.450, -0.514) & 0.099 (-0.521, 0.634)\\
Age 65--100 & Frank & 8.9 & No & -3.731 (-4.792, -2.777) & -1.285 (-1.912, -0.734)\\
Age 65--100 & Gumbel & -17.7 & No & -1.998 (-2.860, -1.241) & -0.391 (-0.924, 0.084)\\
Age 65--100 & Joe & -42.5 & No & -1.048 (-1.839, -0.322) & 0.219 (-0.268, 0.671)\\
Age 65--100 & Survival Clayton & -16.4 & No & -2.106 (-2.989, -1.306) & -0.606 (-1.169, -0.118)\\
ESI 1--2 & Gaussian & -79.5 & No & -3.639 (-4.759, -2.526) & -1.308 (-1.962, -0.699)\\
ESI 1--2 & Student $t$ & -451.2 & No & -0.121 (-1.179, 1.001) & 0.787 (0.167, 1.390)\\
ESI 1--2 & Frank & -152.5 & No & -3.505 (-4.599, -2.295) & -1.394 (-2.067, -0.767)\\
ESI 1--2 & Gumbel & -149.2 & No & -0.729 (-1.681, 0.265) & 0.585 (-0.004, 1.142)\\
ESI 1--2 & Joe & -109.3 & No & -0.751 (-1.554, 0.150) & 0.743 (0.185, 1.274)\\
ESI 1--2 & Survival Clayton & -103.0 & No & -1.877 (-2.855, -0.910) & -0.352 (-0.934, 0.206)\\
ESI 3 & Gaussian & -267.1 & No & -1.090 (-1.859, -0.500) & -0.335 (-0.756, 0.054)\\
ESI 3 & Student $t$ & -922.8 & No & 2.097 (1.386, 2.651) & 1.534 (1.150, 1.936)\\
ESI 3 & Frank & -513.7 & No & -0.898 (-1.671, -0.268) & -0.425 (-0.847, -0.041)\\
ESI 3 & Gumbel & -318.5 & No & 1.375 (0.657, 1.889) & 1.358 (0.986, 1.748)\\
ESI 3 & Joe & -169.8 & No & 0.620 (-0.027, 1.091) & 1.089 (0.758, 1.458)\\
ESI 3 & Survival Clayton & -150.5 & No & -0.300 (-0.985, 0.227) & 0.189 (-0.185, 0.575)\\
Female & Gaussian & -51.4 & No & -3.564 (-4.204, -2.654) & -1.536 (-2.080, -1.106)\\
Female & Student $t$ & -506.1 & No & -0.853 (-1.491, 0.000) & 0.065 (-0.453, 0.472)\\
Female & Frank & -98.4 & No & -3.505 (-4.174, -2.583) & -1.591 (-2.154, -1.155)\\
Female & Gumbel & -175.8 & No & -1.112 (-1.637, -0.363) & -0.015 (-0.489, 0.354)\\
Female & Joe & -163.5 & No & -0.730 (-1.178, -0.026) & 0.337 (-0.102, 0.671)\\
Female & Survival Clayton & -131.9 & No & -1.814 (-2.367, -1.033) & -0.661 (-1.157, -0.295)\\
Male & Gaussian & -92.9 & No & -4.623 (-5.447, -3.868) & -1.544 (-1.965, -1.079)\\
Male & Student $t$ & -997.4 & No & -0.513 (-1.320, 0.205) & 0.814 (0.375, 1.283)\\
Male & Frank & -35.4 & No & -4.014 (-4.849, -3.250) & -1.293 (-1.716, -0.826)\\
Male & Gumbel & 7.8 & No & -2.861 (-3.560, -2.216) & -0.861 (-1.214, -0.413)\\
Male & Joe & -5.8 & No & -2.162 (-2.825, -1.580) & -0.437 (-0.760, -0.006)\\
Male & Survival Clayton & 11.3 & Yes & -3.219 (-4.042, -2.380) & -1.063 (-1.484, -0.572)\\
Overall & BB1 & -818.3 & No & -0.415 (-0.882, 0.004) & 0.542 (0.256, 0.830)\\
Overall & BB7 & -768.8 & No & -0.703 (-1.135, -0.304) & 0.473 (0.203, 0.739)\\
Age 18--44 & BB1 & -857.9 & No & 2.118 (1.296, 2.775) & 1.540 (1.044, 2.014)\\
Age 18--44 & BB7 & -793.2 & No & 1.342 (0.573, 2.004) & 1.253 (0.785, 1.701)\\
Age 45--64 & BB1 & -46.8 & No & -1.686 (-2.577, -0.942) & -0.139 (-0.653, 0.370)\\
Age 45--64 & BB7 & -41.1 & No & -1.799 (-2.640, -1.087) & -0.139 (-0.636, 0.359)\\
Age 65--100 & BB1 & -7.3 & Yes & -1.998 (-2.860, -1.241) & -0.391 (-0.924, 0.084)\\
Age 65--100 & BB7 & -1.6 & Yes & -1.161 (-1.960, -0.429) & 0.128 (-0.359, 0.573)\\
ESI 1--2 & BB1 & -147.8 & No & -1.339 (-2.179, -0.498) & 0.205 (-0.342, 0.724)\\
ESI 1--2 & BB7 & -136.7 & No & -1.449 (-2.258, -0.673) & 0.244 (-0.266, 0.728)\\
ESI 3 & BB1 & -439.5 & No & -0.201 (-0.813, 0.281) & 0.376 (0.054, 0.702)\\
ESI 3 & BB7 & -419.2 & No & -0.511 (-1.140, -0.045) & 0.257 (-0.044, 0.566)\\
Female & BB1 & -164.8 & Yes & -1.112 (-1.637, -0.394) & -0.015 (-0.501, 0.335)\\
Female & BB7 & -158.2 & No & -0.904 (-1.377, -0.232) & 0.211 (-0.245, 0.513)\\
Male & BB1 & 18.6 & Yes & -2.861 (-3.563, -2.216) & -0.861 (-1.214, -0.419)\\
Male & BB7 & 40.7 & Yes & -2.236 (-2.900, -1.649) & -0.507 (-0.829, -0.073)\\
\bottomrule
\end{longtable}
\endgroup

The ratio also depended on which encounters were included. With the first-encounter cut-offs held fixed, including repeat encounters lowered the ratio from 1.63 to 1.26. Using the all-encounter cut-offs lowered it by a further 0.11, to 1.15, and raised the creatinine cut-off from 1.70 to 2.23 mg/dL. Most of the decrease remained when the first-encounter cut-offs were held fixed, so it reflects the inclusion of repeat encounters, and the extra weight they give to frequent visitors, rather than the higher cut-offs. Each visit brings a new triage panel, so a repeat encounter repeats the whole panel rather than rechecking one abnormal value. One possible explanation is that patients with many visits often have chronic conditions in which one measurement stays high and the other does not, for example a raised creatinine from chronic kidney disease with a normal AST. Such patients would add many encounters that count toward one margin only, pulling the ratio toward one. We have not tested this explanation directly, and stratifying by number of encounters would help to test it. The first-encounter estimate therefore describes a patient's presentation more closely, while the all-encounter figure is weighted toward frequent visitors.

\section{Laboratory record selection}\label{app:records}

The analysis used encounter, laboratory and patient-mapping extracts from one institutional clinical warehouse. Records were joined within that source on the encounter identifier. Care-area and emergency admission/discharge/transfer records supplied the encounter-to-patient mapping. A retained encounter required one unique nonblank patient identifier. In the initial eligible encounter pool, fewer than 11 encounters lacked a patient identifier and none had conflicting patient identifiers. No external database was linked. The selection and cleaning rules below use recorded values, units, times and verification status. The study endpoint is the pair of recorded results and does not require a diagnosis-code algorithm.

Within each encounter, specimen and analyte, we collapse identical clinical records and use the latest verification time. The selected record must have Verified (status 9) or Corrected (status 20) status; an older final result is not substituted. If all verification times are missing, we retain a result only when one distinct record remains. Mixed missing and recorded verification times are excluded. Records tied at the latest verification time are merged only when their statuses are accepted, numeric values agree without reporting-limit markers or conflicting value fields, units match exactly and collection times agree; unresolved conflicts are excluded.

We select the earliest structurally eligible specimen pair before applying the numeric exclusions below. Encounters with more than one specimen at that earliest collection time are excluded. The numeric value field is used, with the result field as a fallback only when the value field is blank. The encounter is then excluded if either selected result contains a reporting-limit marker ($<$, $>$, $\leq$ or $\geq$), the two numeric fields disagree, or the selected value is non-numeric, non-finite or not strictly positive. A later specimen is not substituted when the selected pair fails these checks.

\FloatBarrier
\section{Model selection and the paired tail check}\label{app:selection}
\subsection{Data generation and targets}
The unrotated BB1 and BB7 copulas use the parameter order in Appendix~\ref{app:families}. Their overall primary-sample maximum pseudo-likelihood estimates $(\theta,\delta)$ were $(0.04673691,1.03247337)$ for BB1 and $(1.03415177,0.06631905)$ for BB7. The simulation used the full-precision estimates. Independent uniforms followed by conditional inversion generated copula pairs, and inverse normal transformation gave standard normal margins. In two-visit layouts the second pair was copied from the first with probability 0.4. Both coordinates in rounded layouts were rounded to the nearest 0.1. The 107,387-patient layout reproduces the sample size and one-visit design of the application, with simulated normal margins and a common rounding increment.

In each replicate the six main families competed on the same mid-ranks. Each of 999 patient resamples retained a patient's visits together and recomputed cut-offs, ranks, fitted copulas and observed joint rates. The pointwise percentile interval for each family's fitted-minus-observed gap used the 2.5th and 97.5th bootstrap percentiles. The selected-family flag refers to the family chosen in the original replicate. Each bootstrap draw refitted that family. Selection was not repeated within the bootstrap draw.

For a replicate with recorded cut-offs $(x,y)$, let $a$ and $b$ be their known marginal distribution functions evaluated at those cut-offs. A rounded normal margin has $a=\Phi(x+0.05)$ at a recorded value $x$; a continuous normal margin has $a=\Phi(x)$. The generating rate is $1-a-b+C_0(a,b)$ for the generating copula $C_0$. We compared each fitted rate, computed with empirical marginal proportions as in the application, with this known generating rate. The closest family in a replicate minimises the absolute difference. Relative error is the fitted rate minus the generating rate, divided by the generating rate.

The population working parameter maximises expected log copula density at the population mid-distribution ranks. For rounded measurements the calculation used normal-score bins of width 0.1, with open-ended outer bins. For continuous measurements the grid width was 0.01, checked against 0.02. The generating CDF was checked against an independent expression, and two optimisation procedures were compared on the resulting tail probabilities. Population relative errors use the population recorded cut-offs. The replicate-specific gap target for interval coverage substitutes the replicate's cut-offs into both the population working copula and the generating copula, with the true marginal probabilities in both. For the fixed-cut-off target, both copulas are evaluated at the population cut-offs.

\subsection{Selection, errors and false flags}
Table~\ref{tab:selection90} gives the 90th-percentile counterpart to Table~\ref{tab:selection} in the main paper. Table~\ref{tab:largest} compares the six working families in the largest layout. In the rounded population the Student $t$ projection had 30 degrees of freedom under each generator, at the upper fitting limit. At 107,387 observations the main Student $t$ fit reached a parameter bound in 1,731 of 2,000 BB1 replicates and 1,422 of 2,000 BB7 replicates. Selected-family and boundary-fit interval performance is descriptive.

Table~\ref{tab:null} gives false-flag rates for the fixed Gaussian fit under independent margins. With independence the true gap is zero at every cut-off. The denominator for each percentage is 2,000 replicates; Monte Carlo standard errors use the binomial variance divided by that denominator.

\begin{table}[!htbp]
\caption{Selection performance at the 90th percentile.}\label{tab:selection90}
\centering\begingroup\small
\setlength{\tabcolsep}{4pt}
\begin{tabular}{@{}llrrrrr@{}}\toprule
Generator & Layout & Observations & \shortstack{Mean joint\\count} & \shortstack{Winner\\closest (\%)} & \shortstack{Winner\\flagged (\%)} & \shortstack{MCSE for\\flag (pp)}\\ \midrule
BB1 & Continuous & 8,192 & 115.5 & 15.95 & 10.25 & 0.68\\
BB1 & Rounded, two visits & 8,192 & 96.4 & 15.40 & 6.55 & 0.55\\
BB1 & Continuous & 32,768 & 461.0 & 43.80 & 17.55 & 0.85\\
BB1 & Rounded, two visits & 32,768 & 375.4 & 35.10 & 17.25 & 0.84\\
BB1 & Rounded, one visit & 107,387 & 1234.1 & 57.85 & 32.40 & 1.05\\
BB7 & Continuous & 8,192 & 112.7 & 18.10 & 8.85 & 0.64\\
BB7 & Rounded, two visits & 8,192 & 94.3 & 18.15 & 6.00 & 0.53\\
BB7 & Continuous & 32,768 & 451.7 & 54.05 & 11.70 & 0.72\\
BB7 & Rounded, two visits & 32,768 & 368.3 & 43.75 & 10.70 & 0.69\\
BB7 & Rounded, one visit & 107,387 & 1206.7 & 63.10 & 17.65 & 0.85\\
\bottomrule\end{tabular}\endgroup
\par\smallskip\begin{minipage}{0.97\linewidth}\footnotesize The winning family is selected once per replicate, so the selection frequencies are the same at both percentiles. MCSE denotes Monte Carlo standard error and pp denotes percentage points.\end{minipage}
\end{table}
\begin{table}[!htbp]
\caption{Population and sample tail errors and flag rates for 107,387 rounded observations.}\label{tab:largest}
\centering\begingroup\footnotesize
\setlength{\tabcolsep}{4pt}
\begin{tabular}{@{}llrrrrrr@{}}\toprule
 & & \multicolumn{3}{c}{90th percentile (\%)} & \multicolumn{3}{c}{95th percentile (\%)}\\\cmidrule(lr){3-5}\cmidrule(lr){6-8}Generator & Family & Pop. error & Mean error & Flagged & Pop. error & Mean error & Flagged\\ \midrule
BB1 & Gaussian & -10.36 & -10.37 & 99.80 & -23.08 & -24.18 & 100.00\\
BB1 & Student $t$ & -3.41 & -3.37 & 32.20 & -11.29 & -11.97 & 83.65\\
BB1 & Joe & 9.13 & 9.21 & 99.85 & 18.96 & 19.91 & 100.00\\
BB1 & Gumbel & 10.10 & 10.14 & 99.90 & 17.06 & 17.68 & 99.75\\
BB1 & Frank & -17.23 & -17.25 & 100.00 & -33.34 & -34.63 & 100.00\\
BB1 & Survival Clayton & 4.23 & 4.26 & 53.70 & 0.97 & 0.59 & 3.55\\
BB7 & Gaussian & -9.48 & -9.51 & 98.05 & -22.98 & -24.20 & 100.00\\
BB7 & Student $t$ & -2.49 & -2.40 & 17.60 & -11.16 & -11.82 & 82.00\\
BB7 & Joe & 8.26 & 8.25 & 99.55 & 15.86 & 16.50 & 99.80\\
BB7 & Gumbel & 9.88 & 9.85 & 99.85 & 15.10 & 15.49 & 98.50\\
BB7 & Frank & -16.47 & -16.51 & 100.00 & -33.15 & -34.54 & 100.00\\
BB7 & Survival Clayton & 3.66 & 3.62 & 43.15 & -1.21 & -1.83 & 5.90\\
\bottomrule\end{tabular}\endgroup
\par\smallskip\begin{minipage}{0.97\linewidth}\footnotesize Pop. error is the relative error of the population projection at the population cut-offs. Mean error averages the fitted relative errors at the individual replicates' recorded cut-offs. Flagged is the percentage of intervals excluding zero for the named family regardless of selection.\end{minipage}
\end{table}
\begin{table}[!htbp]
\caption{False flags for Gaussian fits to independent measurements.}\label{tab:null}
\centering\begingroup\footnotesize
\setlength{\tabcolsep}{4pt}
\begin{tabular}{@{}lrrrrrrrr@{}}\toprule
 & & & \multicolumn{3}{c}{90th percentile} & \multicolumn{3}{c}{95th percentile}\\\cmidrule(lr){4-6}\cmidrule(lr){7-9}Layout & Obs. & Patients & Count & Flag (\%) & MCSE & Count & Flag (\%) & MCSE\\ \midrule
Continuous & 512 & 512 & 5.10 & 2.80 & 0.37 & 1.22 & 12.05 & 0.73\\
Rounded, two visits & 512 & 256 & 4.24 & 3.65 & 0.42 & 0.98 & 23.30 & 0.95\\
Continuous & 2,048 & 2,048 & 20.35 & 3.45 & 0.41 & 5.13 & 4.45 & 0.46\\
Rounded, two visits & 2,048 & 1,024 & 17.05 & 3.50 & 0.41 & 4.09 & 5.70 & 0.52\\
Continuous & 8,192 & 8,192 & 81.43 & 4.25 & 0.45 & 20.39 & 3.85 & 0.43\\
Rounded, two visits & 8,192 & 4,096 & 66.05 & 4.00 & 0.44 & 16.87 & 4.85 & 0.48\\
Continuous & 32,768 & 32,768 & 327.93 & 4.45 & 0.46 & 81.86 & 5.45 & 0.51\\
Rounded, two visits & 32,768 & 16,384 & 256.99 & 5.30 & 0.50 & 69.66 & 3.95 & 0.44\\
\bottomrule\end{tabular}\endgroup
\par\smallskip\begin{minipage}{0.97\linewidth}\footnotesize Count is the mean number of joint events. MCSEs are in percentage points. The two-visit layout copies the entire pair before rounding.\end{minipage}
\end{table}

\FloatBarrier
\bibliographystyle{unsrtnat}
\bibliography{references}